\documentclass[preprint,12pt]{elsarticle}

\usepackage{amssymb}

\usepackage[numbers]{natbib}
\usepackage{amsmath}
\usepackage{bm}
\usepackage{derivative}
\usepackage{xcolor}
\usepackage{algorithm}
\usepackage{algpseudocode}
\usepackage{subcaption}
\usepackage{tikz}
\usetikzlibrary{decorations.pathreplacing}
\usepackage{hyperref}
\usepackage[left]{lineno}

\algnewcommand\algorithmicobjective{\textbf{Objective:}}
\algnewcommand\Objective{\item[\algorithmicobjective]}

\journal{}

\begin{document}

\begin{frontmatter}



\title{An adaptive, data-driven multiscale approach for dense granular flows}


\author[inst1]{B. Siddani}
\ead{bsiddani@lbl.gov}
\affiliation[inst1]{organization={Center for Computational Sciences and Engineering, Lawrence Berkeley National Laboratory},
            addressline={1 Cyclotron Rd}, 
            city={Berkeley},
            postcode={94720}, 
            state={CA},
            country={United States}}
\author[inst1]{Weiqun Zhang}
\author[inst1]{Andrew Nonaka}
\author[inst1]{John Bell}
\author[inst1]{Ishan Srivastava}
\ead{isriva@lbl.gov}

\begin{abstract}
The accuracy of coarse-grained continuum models of dense granular flows is limited by the lack of high-fidelity closure models for granular rheology.  One approach to addressing this issue, referred to as the hierarchical multiscale method, is to use a high-fidelity fine-grained model to compute the closure terms needed by the coarse-grained model. The difficulty with this approach is that the overall model can become computationally intractable due to the high computational cost of the high-fidelity model. In this work, we describe a multiscale modeling approach for dense granular flows that utilizes neural networks trained using high-fidelity discrete element method (DEM) simulations to approximate the constitutive granular rheology for a continuum incompressible flow model. Our approach leverages an ensemble of neural networks to estimate predictive uncertainty that allows us to determine whether the rheology at a given point is accurately represented by the neural network model. Additional DEM simulations are only performed when needed, minimizing the number of additional DEM simulations required when updating the rheology. This adaptive coupling significantly reduces the overall computational cost of the approach while controlling the error.
In addition, the neural networks are customized to learn regularized rheological behavior to ensure well-posedness of the continuum solution. We first validate the approach using two-dimensional steady-state and decelerating inclined flows. We then demonstrate the efficiency of our approach by modeling three-dimensional sub-aerial granular column collapse for varying initial column aspect ratios, where our multiscale method compares well with the computationally expensive computational fluid dynamics (CFD)-DEM simulation.
\end{abstract}




\end{frontmatter}


\section{Introduction}
\label{sec:intro}
Granular materials have significant applications in industry and nature; yet, a general theory for their dynamics is still an outstanding scientific challenge. Particularly, the flow of granular materials is highly non-Newtonian and can exhibit complex non-equilibrium phenomena such as flow-arrest \cite{srivastava2019flow,srivastava2022flow} and yield transition \cite{clark2018critical}, shear banding \cite{muhlhaus1987thickness}, creep \cite{srivastava2017slow,deshpande2021perpetual}, nonlocal dynamics in confined geometries \cite{kamrin2012nonlocal}, normal stress effects \cite{srivastava2021viscometric,sun2011constitutive}, and pressure-dependent rheology \cite{jop2006constitutive}.

Granular flows can be classified into three distinct regimes, predominantly for convenience: (1) rate-independent quasi-static flows that are modeled using principles of solid mechanics and plasticity \cite{goddard2014continuum}; (2) rate-dependent dense inertial flows that are modeled using the principles of fluid mechanics \cite{kamrin2024advances}; and (3) gas-like collisional flows that are well-analyzed using kinetic theories \cite{jenkins1983theory}. However, the boundary between these regimes is not distinct and there are ongoing efforts to extend theories across regimes \cite{kim2020power,berzi2024granular}. In this work, we focus our attention on quasi-static and dense granular flows where the granular contact lifetimes are relatively long, inertia is important and the material is predominantly dense.

Two classes of numerical methods are particularly well-studied to model granular flows in this regime: (1) the discrete element method (DEM) \cite{cundall1979discrete}, which models individual particles of the material and tracks their trajectory in space and time; and (2) continuum methods such as the mesh-based finite-volume method (FVM) \cite{lagree2011granular,chauchat2014three,rauter2021compressible} and finite element method (FEM) \cite{ionescu2015viscoplastic}, and meshless methods such as the material point method (MPM) \cite{dunatunga2015continuum,haeri2022three,liang2023multiscale} and smooth particle hydrodynamics (SPH) \cite{minatti2015sph,bui2021smoothed}. DEM is particularly useful for capturing particle-scale physics in granular flows and, in principle, provides the highest fidelity representation of granular dynamics. Furthermore, it can be coupled to computational fluid dynamics (CFD) models such as the finite-volume method (FVM) \cite{rauter2021compressible,chauchat2017sedfoam} and the lattice Boltzmann method (LBM) \cite{younes2023lbm} to model the evolution of the solid granular phase in multiphase flows. However, DEM is computationally expensive, particularly for practical applications; for example, just a cubic meter of sand can contain more than $10^{9}$ particles. On the other hand, continuum methods are easily scalable to practical length and time scales, but require constitutive knowledge to close the set of governing equations such as a rheological relationship between the stress and the strain rate.

Significant recent advances have proposed a viscoplastic, scalar $\mu(I)$ inertial rheology for dense granular flows \cite{jop2006constitutive,kamrin2024advances}, where $\mu$ is a dimensionless stress ratio and $I$ is a dimensionless inertial number. The scalar $\mu$ monotonically increases with $I$ and asymptotes to a constant non-zero value as $I\to0$ \cite{da2005rheophysics,srivastava2021viscometric,srivastava2019flow}. The $\mu(I)$ rheology has shown success in predicting velocity profiles in several idealized simple shear configurations \cite{gdr2004dense,kamrin2024advances}, and extensions to it have incorporated additional physics such as anisotropic normal stress effects \cite{srivastava2021viscometric}, Lode-angle dependence \cite{clemmer2021shear,hu2024onsager}, nonlocal dynamics \cite{kamrin2012nonlocal}, and transient dilatancy \cite{barker2017well,dsouza2020non}.

However, the $\mu(I)$ rheology for dense granular flows depends intimately on a multitude of particle-scale properties such as particle shape \cite{salerno2018effect}, size distribution \cite{tripathi2011rheology,polania2025monodisperse}, surface roughness \cite{srivastava2021viscometric,clemmer2021shear}, elasticity \cite{favier2017rheology} and contact restitution \cite{kumaran2009dynamics}, and a general rheological model that accounts for all these properties is currently unavailable. DEM simulations of a representative volume element (RVE) of a granular material (usually significantly smaller than the scale of the application) can be used to calibrate the rheology for a given set of particle-scale parameters \cite{srivastava2021viscometric,clemmer2021shear}. However, even on a reduced RVE scale, these simulations are too expensive to compute the entire rheology, particularly considering the vast design space of particle-scale properties.

Furthermore, even in cases where the underlying rheology is sufficiently parameterized, continuum solutions of dense granular flows are complicated because of the rigid plastic response of the material as $I\to0$, which leads to ill-posedness of the standard $\mu(I)$ rheology in continuum solvers \cite{barker2015well}. The presence of a yield stress as $I\to0$ requires careful treatment such as using appropriate regularization \cite{lagree2011granular}, optimization methods \cite{daviet2016nonsmooth}, or treating the material as an elasto-plastic solid \cite{kamrin2010nonlinear}. The ill-posedness of the local $\mu(I)$ rheology was reported in Barker {\it et al.} \cite{barker2015well} and confirmed by the appearance of instabilities in continuum numerical solutions with the $\mu(I)$ rheology \cite{barker2017partial}. A regularized model of the rheology was proposed that increased the range of inertial numbers where the rheology is well-posed \cite{barker2017partial}.

The aforementioned evidence suggests that high-fidelity modeling of granular flows necessarily requires multiscale coupling between particle-scale DEM and continuum solvers with appropriately built-in regularization for numerical stability. Hierarchical multiscale methods that couple DEM with a continuum solver have shown promise in accurate and efficient modeling of granular materials \cite{andrade2011multiscale,liang2019multiscale,guo2014coupled,liang2023multiscale}. However, despite
being  more efficient than a full DEM-based simulation, they are still computationally expensive since they require the evaluation of the constitutive response using DEM at every time step of the continuum solution.

There is a growing interest in applying machine learning (ML) methods for efficient modeling of granular materials \cite{fransen2025scientificmachinelearninggranular}. Particularly, active learning of constitutive response of granular material and complex fluids within a continuum simulation can improve the efficiency of hierarchical multiscale methods \cite{karapiperis2021data,qu2023deep,zhao2021active}. In this work, we develop an adaptive hierarchical multiscale method that utilizes DEM simulations for \emph{on-the-fly} training and prediction of dense granular rheology needed to advance an FVM-based continuum solver in time. We specifically focus on quasi-steady, dense granular flows whose dynamics are well-represented by the $\mu(I)$ rheology. We use an ensemble of customized neural networks to globally preserve key rheological properties such as monotonicity and convexity of the $\mu(I)$ relationship, along with incorporating regularization for low inertial numbers \cite{barker2017partial} during the adaptive training to ensure numerical stability of the continuum solution. We utilize multilayer perceptron (MLP) as our choice of neural network because of the ease of incorporating these properties in the surrogate models. Unlike traditional hierarchical multiscale methods, granular rheology is not computed for every grid cell but only for those where the uncertainty in the prediction of the rheology exceeds a predetermined threshold. In this manner, the number of DEM simulations is adaptively reduced as the continuum simulation proceeds in time, thus significantly increasing the computational efficiency of the numerical method. The on-the-fly communication between DEM and continuum solvers is efficiently handled using an open-source custom-built multiple-program-multiple-data software that supports GPUs.

The paper is organized as follows. Section \ref{sec:mu_I} gives a brief overview of the regularized $\mu(I)$ rheology for dense granular flows. Section \ref{sec:methodology} describes the governing continuum equations, computational methods to extract the $\mu(I)$ rheology using DEM simulations of a granular RVE, and the neural network (NN) ensemble architecture for adaptive learning of the granular rheology. Section \ref{sec:multiscale_framework} describes the adaptive hierarchical multiscale coupling between the continuum solver and DEM simulations for on-the-fly training of the NN-ensemble. In Section \ref{sec:numerical_examples}, we present numerical examples that demonstrate the efficiency and accuracy of our adaptive hierarchical multiscale method, including validation on $2$D steady and decelerating inclined granular flow problems, and a $3$D demonstration of the accuracy and efficiency on a more complex problem of sub-aerial granular column collapse.
The numerical results are compared against known theoretical result wherever available, and against CFD-DEM simulation for the case of $3$D granular column collapse. Section \ref{sec:conclusions} presents concluding remarks and directions for future work.

\section{Regularized $\mu(I)$ rheology for dense granular flows}
\label{sec:mu_I}
In the dense regime of granular flows, the fluid dynamics can be characterized by the $\mu(I)$ rheology,
where the non-dimensional inertial number, $I$, and stress ratio, $\mu$, are defined as \cite{gdr2004dense,jop2006constitutive}:
\begin{gather}
    \label{eq:inertial_strainrate_connection}
    I = |\bm{D}| d_{p} \sqrt{\frac{\rho_{p}}{p}},\\
    \mu = \frac{|\bm{\tau}|}{p}.
\end{gather}
Here, $|\bm{D}| = \sqrt{\frac{1}{2} \, \bm{D} \colon \! \bm{D}}$ is the magnitude of strain rate, $d_{p}$ is the mean diameter of particles, $\rho_{p}$ is the particle material density, $p$ is pressure, and $|\bm{\tau}| = \sqrt{\frac{1}{2} \, \bm{\tau} \colon \! \bm{\tau}}$ is the magnitude of deviatoric stress. Various functional forms of $\mu(I)$ rheology have been proposed in the literature; in all these forms, the stress ratio asymptotes to a static yield stress ratio $\mu=\mu_s$ as $I\to0$. As $I$ increases, $\mu$ increases monotonically, where a power law is typically used to describe the $\mu(I)$ relationship whose parameters intimately depend on particle characteristics such as interparticle friction \cite{srivastava2021viscometric}, particle shape \cite{salerno2018effect}, and size distribution.

\subsection{Regularization of the $\mu(I)$ rheological model}
\label{sec:partial_reg}
While the granular rheology described above has been used in continuum simulations of dense granular flows \cite{lagree2011granular}, linear stability analysis as well as numerical simulations have shown that the $\mu(I)$ rheology can lead to an ill-posed system of incompressible Navier-Stokes equations at high and low inertial numbers \cite{barker2015well,barker2017partial}.
Particularly, the stability of the $\mu(I)$ rheology can be determined from the following condition \cite{barker2015well,barker2017partial}:
\begin{gather}
    \label{eq:regularization_cond}
    \mathcal{C}(I) = 4 \left(\frac{I \mu^{'}}{\mu}\right)^2 - 4 \left(\frac{I \mu^{'}}{\mu}\right) + \mu^2 \left(1 - \frac{I \mu^{'}}{2\mu}\right)
\end{gather}
where $\mu^{'} = \odv{\mu}{I}$. The $\mu(I)$ rheology is well-posed when $\mathcal{C}(I) \leq 0$ at moderate inertial numbers, but it becomes ill-posed for low and high inertial numbers where $\mathcal{C}(I) > 0$, as was demonstrated by numerical simulations in previous studies \cite{barker2015well,barker2017partial}.

A necessary property from the above condition is that a monotonically increasing $\mu(I)$ function is needed to ensure numerical stability in continuum simulations \cite{barker2017partial}.
In this work, we impose a monotonically-increasing functional form of the rheology in addition to regularization at low inertial numbers in order to ensure numerical stability of the simulations.
Specifically, we consider the following regularized form of the rheology $\mu^{\text{reg}}(I)$ that ensures stability at low inertial numbers \cite{barker2017partial}:
\begin{gather}
    \mu^{\text{reg}}(I) =
    \left\{
	\begin{array}{ll}
		 \sqrt{\frac{\alpha}{\ln\left(\frac{A_{-}}{I}\right)}} & \mbox{if } I \leq I^{N} \\
		\mu(I) & \mbox{if } I > I^{N}
	\end{array}.
    \right.
\label{eq:regularlize_mu_I}
\end{gather}
In this regularized form of dense granular rheology, the standard $\mu(I)$ function modeled using NN-ensembles is used at moderate inertial numbers $I > I^{N}$, whereas a modified functional form is used for low inertial numbers $I \leq I^{N}$. Here $I^{N}$ is the cutoff inertial number, where $A_{-} = I^{N} \exp\left(\frac{\alpha}{\left(\mu(I^{N})\right)^2}\right)$ and we fix $\alpha=1.9$ \cite{barker2017partial}. We obtain $I^{N}$ as the lower root of the equation $\mathcal{C}(I) = 0$ from the unregularized $\mu(I)$ model.
The choice of the coefficient $A_{-}$ ensures that $\mu^{\text{reg}}(I)$ is continuous at $I = I^{N}$.
Furthermore, we also limit the stress ratio $\mu\leq\sqrt{2}$, since the dynamics are always ill-posed at such high inertial numbers \cite{barker2017partial}. For purposes of brevity in the text that follows, we will drop the superscript in $\mu^{\text{reg}}(I)$ such that $\mu(I)$ will denote the regularized rheological model unless noted otherwise.

\section{Methods}
\label{sec:methodology}
In this section, we describe the continuum governing equations for modeling incompressible dense granular flows. We describe the DEM simulations of a granular RVE that are used to extract the granular rheology in our multiscale modeling framework. This is followed by the details of the 
NN-ensemble used for adaptively training and inferring the granular rheology from DEM simulations during the continuum simulation.
\subsection{Continuum modeling}
The first component of our multiscale modeling framework is a continuum model for the dynamics of dense granular flows. In this work, we utilize variable-density, incompressible Navier-Stokes formulation for the continuum modeling of sub-aerial dense granular flows. We consider a two-component model for a mixture of a fluid (air) of density $\rho_f$ and granular material of density $\rho_g$.
The granular material density, which is maintained constant throughout the simulation, is based on a $60 \%$ volume fraction of particles in the fluid.
The governing equations are:
\begin{subequations}
\label{eq:continuum}
\begin{gather}
    \nabla \cdot \bm{u} = 0\\
    \pdv{c}{t} + \nabla \cdot \left(c \bm{u} \right) = 0 \, ; \quad 
    0 \leq c \leq 1 \\
    \rho \left( \pdv{\bm{u}}{t} + \bm{u} \cdot \nabla \bm{u} \right) = -\nabla p + \nabla \cdot \bm{\tau} + \rho \, \bm{g}\\
    \rho  = \left(1-c\right) \rho_f + c \, \rho_g \\
    \bm{\tau} = 2 \, \eta \, \bm{D}; \quad
    \bm{D} = \frac{1}{2} \left( \nabla \bm{u} + {\nabla \bm{u}}^{T}\right) \\ 
    \frac{1}{\eta} = \frac{1-\hat{c}}{\eta_f} + \frac{\hat{c}}{\eta_g}; \quad  \hat{c} = \frac{\rho_g c}{\rho}
\end{gather}
\end{subequations}
where, $\bm{u} = (u,v,w)$ is fluid velocity, $c$ and $\hat{c}$ are the volume and mass fractions of the continuum granular phase respectively, $\rho$ is the
density of the mixture, $p$ is the hydrodynamic pressure, $\bm{\tau}$ is the shear stress tensor, $\bm{g}$ is the gravity vector, $\bm{D}$ is strain rate tensor, and $\eta$ is the dynamic viscosity of the mixture given by the mixture rule\footnote{The sensitivity of the numerical solution to various mixture rules for viscosity is described in \ref{appendix_mixture_viscosity}.} in Eq. \ref{eq:continuum}(f).
The finite-volume numerical solution of the continuum model is based on the approximate projection method for variable density, incompressible Navier-Stokes equations in Almgren {\it et al.} \cite{almgren1998conservative}.
The treatment of variable viscosity uses the methodology in Sverdrup {\it et al.} \cite{sverdrup2018highly, sverdrup2019embedded}.
The Bell, Dawson, and Shubin (BDS) algorithm \cite{bell1988unsplit,nonaka2011three} is used for the advection of the scalar $c$.

The viscosity of the fluid, $\eta_f$, is assumed constant, whereas the viscosity of the granular material, $\eta_g$, is estimated using an NN-ensemble that is trained adaptively by DEM simulations of a granular RVE to compute the $\mu(I)$ rheology. As such, the granular viscosity $\eta_g$ is related to  the $\mu(I)$ rheology by:
\begin{gather}
    \label{eq:gran_visc}
    \eta_g = \frac{\mu(I) \, p_{s}}{2|\bm{D}|},
\end{gather}
where the hydrostatic pressure, $p_s(\bm{r}) = \int \rho \bm{g} \cdot d\bm{r}$, is used to compute both the inertial number in Eq. \eqref{eq:inertial_strainrate_connection}, and the granular viscosity in Eq. \eqref{eq:gran_visc}.

\subsection{Discrete element method (DEM)}
\label{sec:DEMsim}
Here we discuss how discrete element method (DEM) simulations are used to characterize the $\mu(I)$ rheology of granular materials, which is the unknown closure term in the continuum equations. We formulate DEM simulations of an RVE in which a collection of particles, confined within a triclinic periodic cell, are subjected to an external target pressure $p_p^{t}$ and sheared at strain rate $\dot{\gamma} = \frac{I}{d_p}\sqrt{p_p^{t}/\rho_p}$ for a given inertial number $I$, where $\rho_p$ and $d_p$ are respectively the density and average diameter of the particles. The continuum fluid phase is not considered in the DEM simulations that are used to only compute the dry granular rheology. The characteristic time scale in the simulation is the characteristic collision time between two particles $t_c = \pi\left(2k_n/\rho_p d_{p}^{3} - \nu_n^2/4\right)^{-1/2}$, where $k_n$ and $\nu_n$ are respectively the normal Hookean spring constant and normal damping coefficient of contacting particles \cite{silbert2001granular}. The stiffness $k_n$ sets the stress scale in the system, and all stresses are scaled by $k_n/d_p$.

The system is initialized in a cubic periodic cell below jamming at a volume fraction $\phi_0$. The triclinic cell is first subjected to only compression
with a time step of $0.2\sqrt{\rho_{p}d_{p}^{3}/k_{n}}$ to achieve the target pressure, $p_{p}^{t}$. The periodic cell is then deformed at a constant strain rate $\dot{\gamma}$ by affinely remapping the position of all particles at each time step followed by a velocity-Verlet integration of their forces, velocities and positions. As large strains are accumulated in long simulations, the sheared boundaries are flipped periodically to prevent an extremely skewed periodic cell. The pressure in the system, $p_p$, is maintained at its target value of $p_p^{t}$,
which is usually chosen small enough to correspond to the hard-particle limit that is appropriate for typical granular material. The pressure is maintained by isotropically expanding or contracting the periodic cell whose cell length $L(t)$ is linearly controlled by:
\begin{equation}
    \frac{\text{d}L(t)}{\text{d}t} = L(t)k_V\left(p_p^{t} - p_p\right),
\end{equation}
where $k_V$ is a proportional gain constant whose magnitude controls the balance between volumetric and pressure fluctuations; we use a value of $k_V=10^{-4}$ in all the simulations.

In the DEM simulations, all components of the total stress tensor $\bm{\sigma}_{\text{rve}}$ of the RVE are calculated as \cite{walton1986viscosity,da2005rheophysics}:
\begin{equation}
\sigma_{\alpha\beta,\text{rve}}=\frac{1}{V_{\text{rve}}}\sum_{i}\left[\sum_{j \neq i}\frac{1}{2}r_{\alpha,ij} f_{\beta,ij} + m_{i}v_{\alpha,i}v_{\beta,i}\right],
\end{equation}
where for each pair of contacting particles $i$ and $j$, $\boldsymbol{r}_{ij}$ and $\boldsymbol{f}_{ij}$ are the interparticle branch vector and force vector, and $\boldsymbol{v}_{i}$ and $\boldsymbol{v}_{j}$ are the particle velocities. The first term corresponds to a `virial' stress resulting from interparticle contacts, and the second term results from the momentum of the particles. Here $V_{\text{rve}}$ is the volume of the periodic cell. The hydrostatic pressure is computed as $p_p=(1/3)\sum_i \sigma_{ii,\text{rve}}$ and shear stress is computed as $|\bm{\tau}_{\text{rve}}| = \sqrt{\frac{1}{2} \, \bm{\tau}_{\text{rve}} \colon \! \bm{\tau}_{\text{rve}}}$, where the deviatoric stress $\bm{\tau}_{\text{rve}} = \bm{\sigma}_{\text{rve}} - p_p\mathbb{I}$, and $\mathbb{I}$ is the identity tensor. Subsequently, the stress ratio of the RVE is computed as:
\begin{gather}
    \mu=|\bm{\tau}_{\text{rve}}|/p_p.
\end{gather} 

\subsection{Neural network ensemble for learning regularized $\mu(I)$ rheology}
The last component of our multiscale modeling framework is the NN-ensemble that is utilized for active learning of the $\mu(I)$ rheology from DEM simulations {\it in-situ}  during a continuum simulation. For numerical stability of the continuum simulations, we incorporate monotonicity and convexity of $\mu$ with respect to $\ln(I)$ into the MLP \cite{liu2020certified}.
The architecture of a $k-$layer MLP used in the current work is given by
\begin{gather*}
    \bm{h}_{i+1} = \mathcal{F}_{i}(\bm{W}_{i} \bm{h}_{i} + \bm{b}_{i}), \, \text{for} \, i=0,1,...,k-1
\end{gather*}
where $\bm{h}_{0} = \ln(I)$ and $\bm{h}_{k} = \mu$ are input and output of the MLP respectively, $\bm{\theta} = \{\bm{W}_{0:k-1}, \bm{b}_{0:k-1}\}$ are parameters of the MLP, and $\mathcal{F}_{0:k-1}$ are activation functions. The desired properties require $\bm{W}_{0:k-1}$ are non-negative and $\mathcal{F}_{0:k-1}$ are convex and non-decreasing functions. We leverage ReLU (Rectified Linear Unit) activation function for $\mathcal{F}_{0:k-2}$ and the identity function for $\mathcal{F}_{k-1}$. There exist NN architectures that impose only convex property with respect to some or all of the inputs \cite{amos2017input}.

Instead of a point estimate associated with a single MLP, we use an ensemble of MLPs \cite{lakshminarayanan2017simple} to obtain the uncertainty associated with the NN approximation of the $\mu(I)$ rheology.
The tunable parameters of each MLP in the ensemble are initialized to different values such that they are optimized to different parameter sets during the training process. As such, the NN-ensemble will produce a distribution of $\mu$ for a given $I$.
The NN-ensemble of size $M$ is denoted as $\{\mu_{m}(I)\}_{m=1}^{M}$, where the output of the $m$-th MLP is $\mu_{m}$ for a given input, which is the logarithm of the inertial number, $\ln(I)$. The loss function that is minimized to train the NN-ensemble is:
\begin{gather}
    \min\limits_{\{\bm{\theta}_{m}\}_{m=1}^{M}} \mathcal{L}_{\text{MSE}}(\bm{I}_{\text{train}}) - \lambda \mathcal{L}_{\text{MOD}}(\bm{I}_{\text{mod}}),
\end{gather}
where,
\begin{gather}
    \mathcal{L}_{\text{MSE}}(\bm{I}_{\text{train}}) = \frac{1}{M} \sum\limits_{m=1}^{M} \frac{1}{N_{\text{train}}} \sum\limits_{I \in \bm{I}_{\text{train}}} \left[\mu_{m}(I) - (\Tilde{\mu}(I) + \sigma_{\Tilde{\mu}(I)} \epsilon)\right]^2, \\
    \mathcal{L}_{\text{MOD}}(\bm{I}_{\text{mod}}) = \frac{1}{N_{\text{mod}} (M-1)} \sum\limits_{I \in \bm{I}_{\text{mod}}} \sum_{m=1}^{M}\left(\mu_{m}(I) - \sum_{m'=1}^{M} \mu_{m'}(I)/M\right)^2.
\end{gather}
Here $\bm{\theta}_{m}$ are weights and biases of the $m$-th MLP, $\bm{I}_{\text{train}}$ is the ordered (in ascending value) training dataset of size $N_{\text{train}}$, and $\Tilde{\mu}(I)$ and $\sigma_{\Tilde{\mu}}(I)$ are the mean and standard error of the stress ratio obtained from DEM simulations for $I \in \bm{I}_{\text{train}}$. $\mathcal{L}_{\text{MSE}}(\bm{I}_{\text{train}})$ is the mean squared error loss for training data, and $\epsilon$, which is unique for each $I \in \bm{I}_{\text{train}}$ and each $m \in \{\mu_{m}\}_{m=1}^{M}$, is sampled from standard normal distribution, i.e., $\epsilon \sim \mathcal{N}(0,1)$. We utilize a `maximize overall diversity' (MOD) technique to improve the uncertainty estimates at non-training points \cite{jain2020maximizing}.
This is accomplished by adding an additional term $\mathcal{L}_{\text{MOD}}$ to the loss function. This term is computed at a collection of points denoted by $\bm{I}_{\text{mod}}$ of size $N_{\text{mod}}$ that are obtained by randomly choosing $10$ values of inertial number in the range $\left[I_{\text{train}}^{(i)} + 0.2(I_{\text{train}}^{(i+1)}-I_{\text{train}}^{(i)}), I_{\text{train}}^{(i)} + 0.8(I_{\text{train}}^{(i+1)}-I_{\text{train}}^{(i)}) \right]$ for $1\leq i < N_{\text{train}}$. 
During minimization of the loss function of the NN-ensemble, the additional term that is included in the loss function with a negative sign maximizes the variance of NN-ensemble predictions at non-training points as it attempts to fit the DEM data at the training points. A small positive value of $\lambda = 3 \times 10^{-3}$ is chosen for all the numerical examples presented here.

The parameters of the NN-ensemble are optimized using the Adam optimizer \cite{kingma2014adam} with a learning rate of $1 \times 10^{-3}$. The training process for a given set $\bm{I}_{\text{train}}$ is conducted over $10^4$ iterations. In this study, we choose an ensemble of MLPs of size $M=10$. The MLP architecture consists of $2$ hidden layers, each containing $100$ neurons.

\section{Adaptive multiscale modeling for dense granular flows}
\label{sec:multiscale_framework}
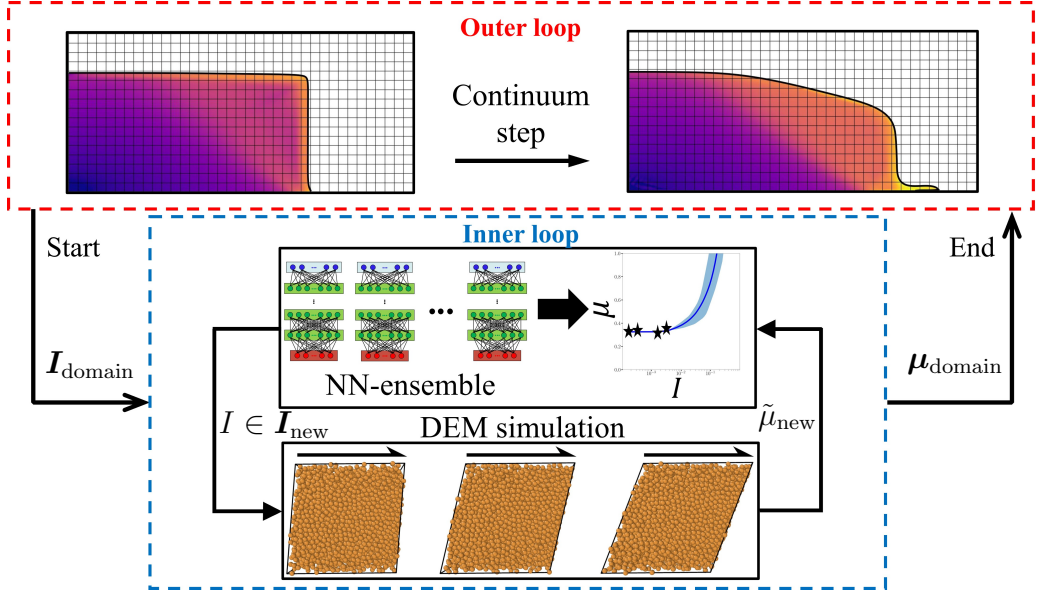
\begin{figure}[htbp]
    \centering
    \includegraphics[width=\linewidth]{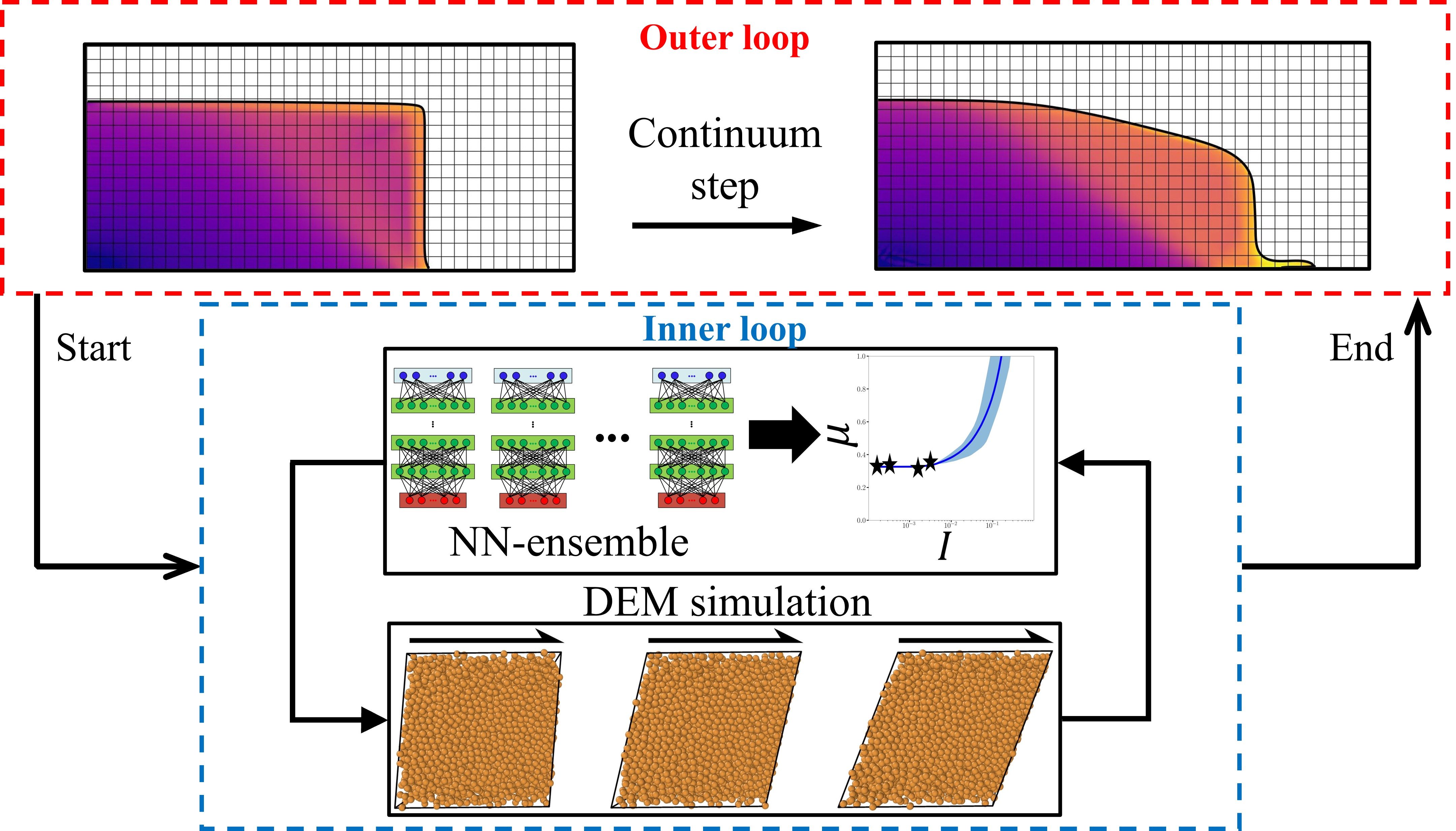}
    \begin{tikzpicture}[overlay, remember picture]
        \node at (-5.75, 3.5) {$\bm{I}_{\text{domain}}$};
        \node at (-3.3, 2.7) {$I \in \bm{I}_{\text{new}}$};
        \node at (3.52, 2.8) {$\Tilde{\mu}_{\text{new}}$};
        \node at (5.75, 3.5) {$\bm{\mu}_{\text{domain}}$};
    \end{tikzpicture}
    \caption{Adaptive hierarchical multiscale method for simulating dense granular flows. The outer loop of the method evolves a coarse-grained continuum incompressible flow simulation, whereas the inner loop of the method combines NN-ensemble and fine-grained DEM simulations to adaptively compute the dense granular rheology. During a continuum time step, the inertial numbers at all finite-volume grid cells, $\bm{I}_{\text{domain}}$, are communicated to the NN-ensemble; inertial numbers with high uncertainty in the NN-ensemble, $\bm{I}_{\text{new}}$, are identified; new DEM simulation is run for the $I \in \bm{I}_{\text{new}}$ with the highest uncertainty to compute $\Tilde{\mu}_{\text{new}}$, followed by updating the NN-ensemble and $\bm{I}_{\text{new}}$. This process is repeated iteratively until $\bm{I}_{\text{new}}$ is empty, upon which the stress ratio set for the entire domain, $\bm{\mu}_{\text{domain}}$, computed from the NN-ensemble is communicated back to the outer loop simulation.}
    \label{fig:multiscale_schematic}
\end{figure}

In this section we describe how our multiscale modeling framework and workflow is constructed using the model components described in the previous section. Each simulation begins with a pre-trained NN ensemble for the $\mu(I)$ rheology where the initial training data can be obtained either from experiments or DEM simulations. In this study, we pre-train our ensemble of NNs with $\mu$ obtained from DEM simulations corresponding to an initial set of inertial numbers $\bm{I}_{0}$. While the choice of $\bm{I}_{0}$ is arbitrary, we choose $\bm{I}_{0}$ clustered at small inertial numbers since they incur the highest computational costs in a DEM simulation. At the beginning of the multiscale simulation, $\bm{I}_{0}$ constitutes the initial training dataset $\bm{I}_{\text{train}}$.

Figure \ref{fig:multiscale_schematic} shows a schematic of our modeling framework and the interactions between various model components during a single time step of a continuum simulation that are described below. Algorithm \ref{alg:multiscale_workflow} in \ref{appendix_multiscale} describes the corresponding algorithmic details of this framework. At the beginning of each continuum step, the continuum solver computes the local inertial number from the local strain rate and hydrostatic pressure at each grid cell where the volume fraction of the granular material $c \geq 1\times 10^{-3}$. This set of inertial numbers $\bm{I}_{\text{domain}}$ is communicated to the NN-ensemble, and evaluated to identify a subset of inertial numbers that require new DEM simulations. The evaluation consists of three stages detailed below. In each of these stages, new DEM simulations may be performed, and the training data, $\bm{I}_{\text{train}}$ and NN-ensemble are updated after each new DEM simulation.

In the first stage, we update $\bm{I}_{\text{train}}$ and the NN-ensemble so that every $I$ in $\bm{I}_{\text{domain}}$ lies between $\min(\bm{I}_{\text{train}})$ and $\max(\bm{I}_{\text{train}})$. Specifically, if $\min(\bm{I}_{\text{domain}}) < \min(\bm{I}_{\text{train}})$, we perform a new DEM simulation for minimum inertial number $\min(\bm{I}_{\text{domain}})$ and update the $\bm{I}_{\text{train}}$ and the NN-ensemble. Similarly, if $\max(\bm{I}_{\text{domain}}) > \max(\bm{I}_{\text{train}})$, we perform a new DEM simulation for maximum inertial number $\max(\bm{I}_{\text{domain}})$ and update the $\bm{I}_{\text{train}}$ and the NN-ensemble.

In the second stage, we estimate the uncertainty $\mathcal{R}(I)$ for every $I \in \bm{I}_{\text{domain}}$ using the NN-ensemble of $M$ neural networks, $\{\mu_{m}\}_{m=1}^{M}$, as:
\begin{gather}
    \mathcal{R}(I) = \text{std}(\{\mu_{m}(I)\}_{m=1}^{M}) /\text{mean}(\{\mu_{m}(I)\}_{m=1}^{M}),
\end{gather}
where
\begin{gather*}
    \text{std}(\{\mu_{m}(I)\}_{m=1}^{M}) = \sqrt{\frac{1}{M-1}\sum_{m=1}^{M}\left(\mu_{m}(I) - \sum_{m'=1}^{M} \mu_{m'}(I)/M\right)^2}, \\
    \text{mean}(\{\mu_{m}(I)\}_{m=1}^{M}) = \frac{1}{M} \sum_{m=1}^{M} \mu_{m}(I) .
\end{gather*}
If $\bm{I}_{\text{domain}}$ contains points where the uncertainty is too high, we perform additional DEM simulations to reduce the uncertainty at those points.  We use an iterative process in which we identify the point with the highest uncertainty, add that point to the training data and perform a new DEM simulation and update the NN-ensemble.  We consider the uncertainty at $I$ as too high when $\mathcal{R}(I)$ exceeds a local threshold uncertainty $\mathcal{T}_{\text{UQ}}$ computed from two nearest training data points, $I_{\text{train}}^{(i)}$ and $I_{\text{train}}^{(i+1)}$, where $I_{\text{train}}^{(i)} < I < I_{\text{train}}^{(i+1)}$. The local threshold uncertainty is defined as:
\begin{gather}
     \mathcal{T}_{\text{UQ}}(I, I_{\text{train}}^{(i)}, I_{\text{train}}^{(i+1)})= \kappa \left[(1-\beta) \mathcal{R}(I_{\text{train}}^{(i)})+ \beta \mathcal{R}(I_{\text{train}}^{(i+1)})\right]
     \label{eq:threshold_uq}
\end{gather}
where
\begin{gather*}
    \beta = \frac{\ln(I) - \ln(I_{\text{train}}^{(i)})}{\ln(I_{\text{train}}^{(i+1)}) - \ln(I_{\text{train}}^{(i)})},
\end{gather*}
and $\kappa$ is a sampling factor that controls the acceptable threshold of uncertainty. When $\mathcal{R}(I) > \mathcal{T}_{\text{UQ}}(I, I_{\text{train}}^{(i)},I_{\text{train}}^{(i+1)})$ for an $I \in \bm{I}_{\text{domain}}$ , that $I$ is identified as a candidate for a new DEM simulation. We then identify the inertial number $I$ corresponding to the highest uncertainty $\mathcal{R}(I)$ and perform a DEM simulation at that point. Then, $\bm{I}_{\text{train}}$ and NN-ensemble are updated, and $\bm{I}_{\text{new}}$ is re-evaluated using the updated NN-ensemble and the threshold uncertainty defined in Eq.~\eqref{eq:threshold_uq}. This operation is conducted iteratively until the uncertainty is below the threshold for all $I\in \bm{I}_{\text{domain}}$. We note that the number of potential additional DEM simulations is controlled by $\kappa$.  A higher value of $\kappa$ allows for higher uncertainty in the model prediction thus requiring fewer DEM simulations.

In the last stage, we regularize the rheology predicted by the NN-ensemble for low inertial numbers, as described in Sec.~\ref{sec:partial_reg}. This process requires identifying the low-cutoff inertial number $I^{N}$ such that for all $I \in \bm{I}_{\text{domain}}$ where $I \leq I^{N}$, the modified functional form of the rheology in Eq.~\eqref{eq:regularlize_mu_I} is used instead of the prediction of the rheology from the NN-ensemble model. We begin by computing the stability condition, $\mathcal{C}(I)$, (see Eq.~\eqref{eq:regularization_cond}) on all the inertial numbers in $\bm{I}_{\text{train}}$. Since the computation of $\mathcal{C}(I)$ requires estimating $\mu^{'} = \odv{\mu}{I}$ from $\mu(I)$ model of the NN-ensemble, we use automatic differentiation for this purpose. If the lower root of $\mathcal{C}(I)=0$ exists within the range of the $\bm{I}_{\text{train}}$, we use an iterative method to identify $I^{N}$ within the range of $\bm{I}_{\text{train}}$. In the present work, we use a total of five iterations to obtain the numerical estimate of $I^{N}$, and the reader is referred to Algorithm~\ref{alg:neutral_num} in \ref{appendix_multiscale} for algorithmic details. In the case where the lower root of $\mathcal{C}(I)=0$ does not exist within the range of the $\bm{I}_{\text{train}}$, we sufficiently expand $\bm{I}_{\text{train}}$ such that it contains the lower root of $\mathcal{C}(I)=0$. For example, if $\mathcal{C}(I) > 0$ for all the inertial numbers in $\bm{I}_{\text{train}}$, we iteratively increase $\max(\bm{I}_{\text{train}})$ by a factor of two until there are at least two points in $\bm{I}_{\text{train}}$ that now satisfy $\mathcal{C}(I) \leq 0$. Each such iteration involves running a DEM simulation for the new $\max(\bm{I}_{\text{train}})$, updating $\bm{I}_{\text{train}}$ and the NN-ensemble, and re-evaluating $\mathcal{C}(I)$ for all the inertial numbers in $\bm{I}_{\text{train}}$ using automatic differentiation. Once the range of $\bm{I}_{\text{train}}$ is sufficiently expanded, the iterative method described above is used to identify $I^{N}$ corresponding to the lower root of $\mathcal{C}(I)=0$ for the updated training data. The reader is referred to Algorithms \ref{alg:new_train_points} and ~\ref{alg:initial_neutral_num} in \ref{appendix_multiscale} for algorithmic details regarding updating the $\bm{I}_{\text{train}}$ using DEM simulations in the evaluation process.

Once $I^{N}$ is suitably determined, the regularized stress ratio $\bm{\mu}_{\text{domain}}$ corresponding to $\bm{I}_{\text{domain}}$ is computed and communicated back to the continuum solver such that granular viscosity can be calculated in each grid cell containing granular material in order to advance the time step of the continuum solution. Here, for all $I \in \bm{I}_{\text{domain}}$ where $I>I^N$, the stress ratio corresponding to the NN-ensemble mean, $\sum_{m=1}^{M} \mu_{m}/M$, is used, whereas for $I\leq I^N$, the regularized functional form of the rheology in Eq.~\eqref{eq:regularlize_mu_I} is used in the computation of $\bm{\mu}_{\text{domain}}$.

\section{Numerical Examples}
\label{sec:numerical_examples}
In this section we first describe the numerical details of the DEM simulations that are used to pre-train and adaptively train the dense granular rheology. Next, we describe three numerical examples of dense granular flows that are simulated using the adaptive multiscale modeling framework discussed above. The first example validates the methodology by simulating a steady-state, two-dimensional inclined chute flow where the velocity profile for a given $\mu(I)$ rheology is known analytically. The second example simulates a two-dimensional decelerating inclined chute flow to demonstrate the numerical stability that is achieved by incorporating low inertial number regularization in the NN-ensemble. Finally, the last example simulates the complex dynamics of three-dimensional sub-aerial granular column collapse to demonstrate the efficacy and accuracy of our adaptive hierarchical multiscale method.

\subsection{Numerical details of DEM simulations}
All DEM simulations comprise of a system of $\sim 10^4$ nearly-monodisperse, spherical particles of unity density $\rho_p$ whose diameters are uniformly distributed between $0.95d_p$ and $1.05d_p$, where $d_p$ is also set to unity. The particles interact on contact through a linear spring-dashpot viscoelastic mechanical model, along with tangential Coulomb friction that is characterized by a coefficient of friction, $\mu_p$ \cite{silbert2001granular}.
Both normal and tangential stiffness are set equal to unity. The normal and tangential damping coefficients are set to $\nu_n=0.5$ and $\nu_t=0.25$ respectively, which corresponds to a coefficient of restitution between $0.62$ and $0.75$ depending on the particle radius. All the
rheological simulations are performed using the granular package in the large-scale
molecular dynamics software LAMMPS \cite{thompson2022lammps}.

Each DEM simulation is initialized at a volume fraction $\phi_0 \sim 0.35$ and then subjected to a compression for $10^5$ steps to achieve a target pressure $p_p^{t}=10^{-5}k_n/d_p$. Thereafter, upon turning on shear deformation as described in Sec.~\ref{sec:DEMsim}, the granular RVE evolves from a transient towards a nonequilibrium steady state where the value of $p_p$ and $\mu$ fluctuate around a mean value. After the initial compression, each DEM simulation is sheared either to a strain of $0.5$ or for a duration of $10^{5}\sqrt{\rho_{p}d_{p}^{3}/k_{n}}$, whichever is longer, using a time step of $0.2\sqrt{\rho_{p}d_{p}^{3}/k_{n}}$. The system is sheared further to a strain of $1.0$ with a reduced time step of $0.02\sqrt{\rho_{p}d_{p}^{3}/k_{n}}$. Upon deforming up to a unity strain, the stress ratio $\mu$ is collected at every $10$ steps for a total of $2\times10^5$ steps. The statistical uncertainty associated with the estimation of mean stress ratio, $\Tilde{\mu}$, from the DEM simulations is characterized by its standard error, $\sigma_{\Tilde{\mu}}$, using a block averaging method \cite{flyvbjerg1989error}.

We restrict conducting DEM simulations to compute $\mu$ for the training data between a minimum $I_{\text{DEM}}^{\min}$ and a maximum $I_{\text{DEM}}^{\max}$ inertial number.
At very low inertial numbers, it is computationally prohibitive to reach large strains and extract statistically significant stresses from the granular RVE in DEM simulations. At very high inertial numbers, the DEM simulations become unstable as the dense granular flow transitions into a gas-like regime. In this work, we set $I_{\text{DEM}}^{\min}=1.6\times10^{-4}$ and $I_{\text{DEM}}^{\max}=3.2\times10^{-1}$ for all the numerical examples below.

\subsection{Inclined chute flow}
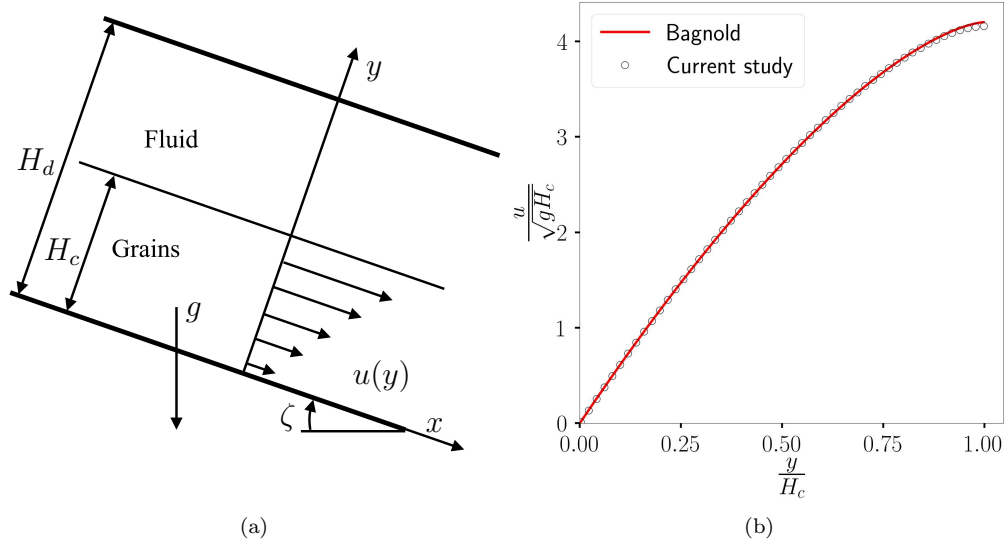
\begin{figure}[htbp]
    \centering
    \begin{subfigure}[b]{0.48\textwidth}
        \centering
        \includegraphics[width=\linewidth,keepaspectratio=true]{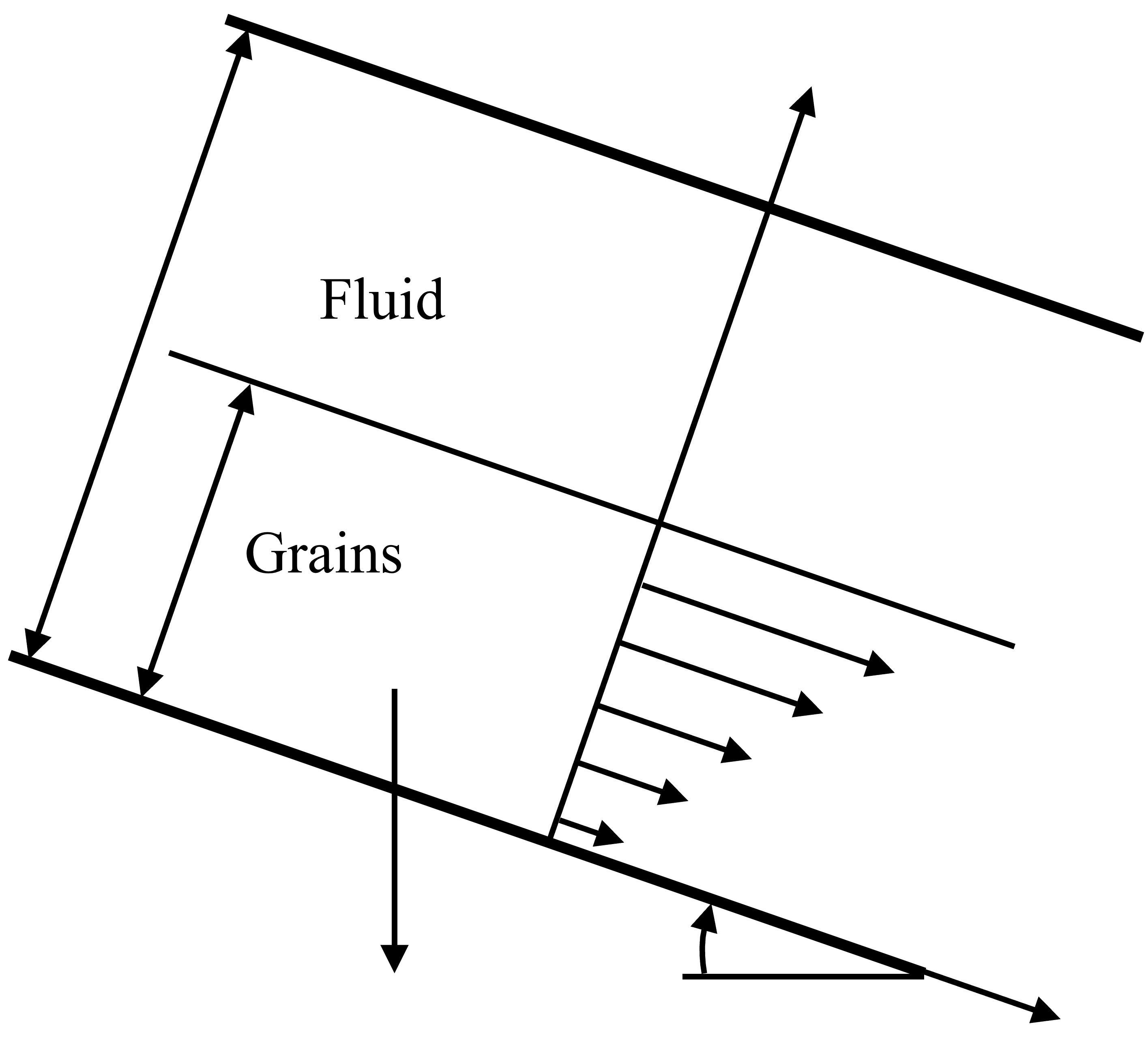}
        \begin{tikzpicture}[overlay, remember picture]
            \node at (-0.8, 2.5) {$g$};
            \node at (-2.55, 3.3) {$H_{c}$};
            \node at (-2.9, 4.5) {$H_{d}$};
            \node at (0.45, 1.1) {$\zeta$};
            \node at (1.7, 1.65) {$u(y)$};
            \node at (1.6, 5.7) {$y$};
            \node at (2.4, 1.0) {$x$};
        \end{tikzpicture}
        \caption{}
    \end{subfigure}
    \begin{subfigure}[b]{0.48\textwidth}
        \centering
        \includegraphics[width=\linewidth,keepaspectratio=true]{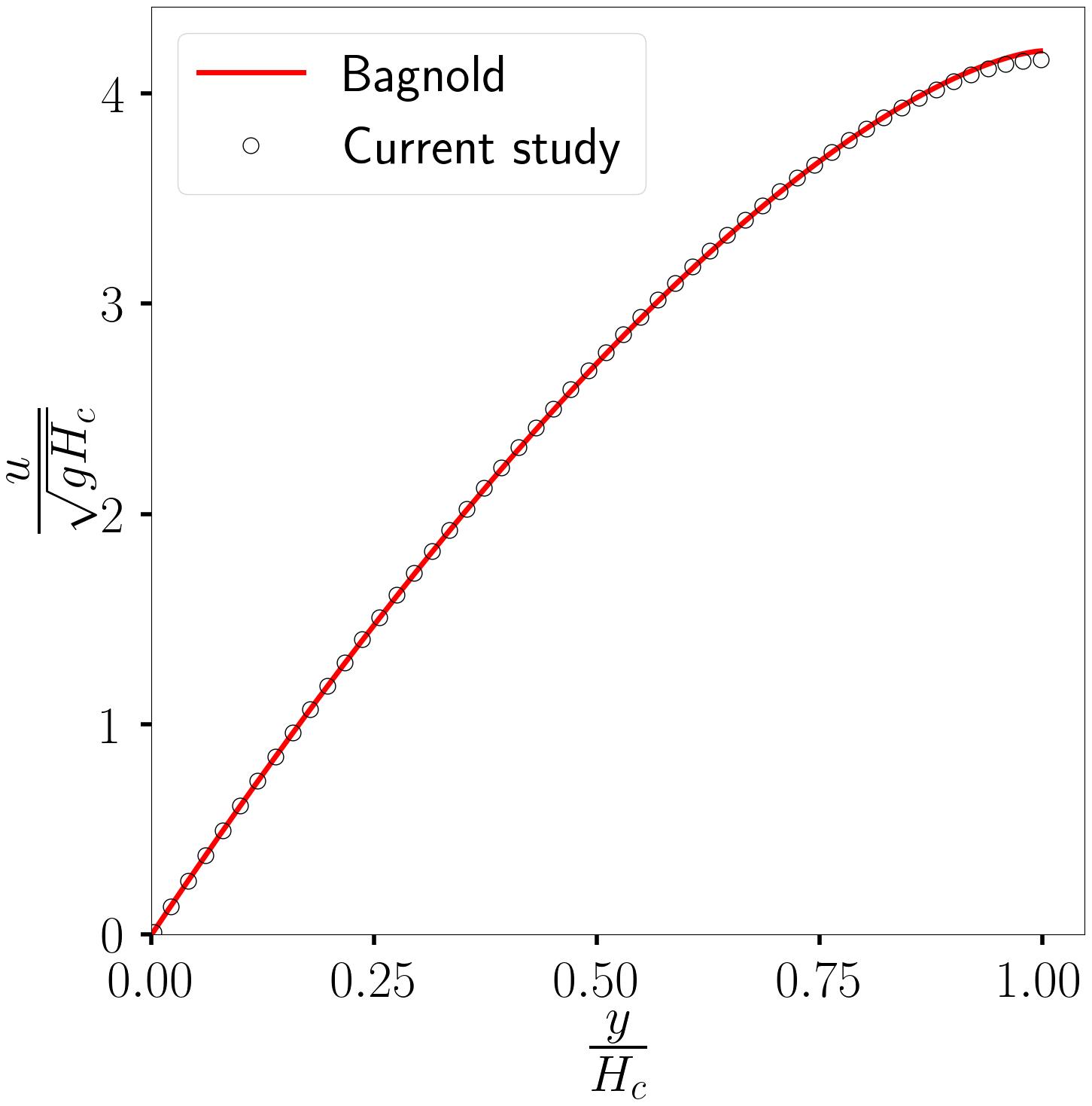}
        \caption{}
    \end{subfigure}
    \caption{(a) Two-layer inclined plane setup. The plane is inclined at an angle $\zeta$ with respect to the horizontal. The domain is initialized with granular material up to height $H_{c}$ and the rest of the domain up to $H_{d}$ is occupied by a low-density Newtonian fluid, (b) Steady state velocity profile in the granular material region, $0 \leq y \leq H_c$, at $t \approx 4$s for $\zeta = 0.43$. The corresponding Bagnold's solution of the velocity profile from Eq. \eqref{eq:bagnold_profile} is also shown.}
    \label{fig:double_layer_bagnold}
\end{figure}

We first simulate a two-dimensional granular flow along an inclined plane, as shown in the schematic in Fig. \ref{fig:double_layer_bagnold}(a). The setup consists of granular material of height $H_c$ ranging from $y=0$ to $y=H_d/2$ and a low-density Newtonian fluid from $y=H_d/2$ to $y=H_d$ initialized on a plane inclined at an angle $\zeta$. No-slip boundary conditions are imposed at $y=0$ and $y=H_d$, while the hydrostatic pressure is set to $0$ at $y=H_d$. The system is periodic in the $x$ direction so that only the $x$-component of the velocity $u$ varies along $y$ direction and is invariant in the $x$ direction. The material parameters and the simulation setup are tabulated in Table \ref{tab:inclined_plane}. The continuum simulation uses a time step such that the advective Courant–Friedrichs–Lewy (CFL) number is $0.35$.

The steady-state solution for the inclined plane flow problem follows the well-known Bagnold's profile \cite{bagnold1954experiments,silbert2001granular}, where $u$ depends on the $\mu(I)$ rheology as:
\begin{gather}
     \label{eq:bagnold_profile}
     u(y) = \frac{4 I_{\zeta}}{3 d_{p}} \sqrt{\frac{\rho_g}{\rho_{p}} g \cos{\zeta}} \left( H_{c}^{3/2} - {(H_{c}-y)}^{3/2}\right)\\
     \mu = \tan{\zeta}; \quad I_{\zeta} = \mu^{-1}(\tan \zeta), \nonumber 
\end{gather}
where $\mu^{-1}$ is the inverse function relating the inertial number to a stress ratio, and $g=9.81\text{m}/\text{s}^2$ is the acceleration due to gravity. Here we consider the plane inclined at an angle $\zeta=0.43$, or $24.6^{\text{o}}$, that is greater than the angle $\text{arctan}(\mu_s)=18.3^{\text{o}}$ corresponding to the static ratio $\mu_s\approx 0.33$ for particles with interparticle friction $\mu_p=0.3$ \cite{srivastava2021viscometric,clemmer2021shear}. The inertial number corresponding to $\mu = \text{tan}(\zeta)$ is estimated as $I_{\zeta}\approx0.11$ \cite{srivastava2021viscometric,clemmer2021shear}.

To quickly reach steady state, the system is initialized with a Bagnold's velocity profile for the granular material corresponding to $\zeta=0.43$, or $24.6^{\text{o}}$, along with a linear $u(y)$ for the lighter fluid such that the velocity is continuous across the granular material-fluid interface. In addition to the regularization of the $\mu(I)$ rheology discussed above, we also set a minimum viscosity for the granular material $\eta_g^{\text{min}} = 0.01\rho_g\sqrt{g{d}_{p}^{3}}$, which corresponds to the smallest value of viscosity at a depth of one particle below the surface \cite{lagree2011granular}. Figure \ref{fig:double_layer_bagnold}(b) shows the normalized velocity profile of the granular material after $t \approx 4$s of simulation that corresponds to $2\times10^5$ steps. The velocity is in excellent agreement with the expected Bagnold's profile. Slight deviation from the Bagnold's profile is observed near the granular material-fluid interface, which is consistent with previous simluations on a similar setup \cite{lagree2011granular}.

This example validates our implementation of dense granular rheology using NN-ensembles in an adaptive hierarchical multiscale method. Using a value of $\kappa=5$, a total of $11$ new DEM simulations in addition to the $4$ initial data points in $\bm{I}_{0}$ were run during the comtinuum simluations. The total number of DEM simulations queried in our adaptive hierarchical multiscale method is significantly lesser than that in direct hierarchical multiscale method, which would require running DEM simulations to compute the rheology at each grid cell for each time step. Furthermore, the current approach also used much fewer DEM simulations than needed to reliably pre-compute the entire $\mu(I)$ across all inertial numbers \cite{srivastava2021viscometric}.

\begin{table}[htbp]
    \centering
    \begin{tabular}{|c|c|}
        \hline
        Parameter & Value\\
        \hline
        $(L_{d}, H_{d})$  & (0.08m, 0.08m) \\
        $\Delta x$ & $1.5625 \times 10^{-4}$m \\
        $\rho_f$ & $1$kg/m$\textsuperscript{3}$\\
        $\rho_p$ & $2600$kg/m$\textsuperscript{3}$\\
        $\rho_g$ & $1560.4$kg/m$\textsuperscript{3}$\\
        $d_p$ & $10^{-3}$m\\
        $\mu_p$ & $0.3$\\
        $\eta_f$ & $2.5\times10^{-4}$Pa$\cdot$s\\
        $\eta_g^{\text{min}}$ & $0.01\rho_g\sqrt{g{d}_{p}^{3}}$\\
        \hline
    \end{tabular}
    \caption{Parameters and material properties used to simulate the two-layer inclined plane granular flows using the multiscale approach.}
    \label{tab:inclined_plane}
\end{table}

\subsection{Decelerating chute flow}
Using the setup described in the previous section, we now consider the more challenging case of decelerating chute flows. The simulation is initialized with the steady-state Bagnold's velocity profile at the angle $\zeta_0=0.43$, or $24.6^{\text{o}}$, for the granular material, and the inclined angle is instantly reduced to $\zeta=0.175$, or $10^{\text{o}}$, which is considerably lower than the dynamic angle of repose $\text{arctan}(\mu_s)=18.3^{\text{o}}$ for the system considered here \cite{srivastava2021viscometric}. In this case, the system is expected to quickly decelerate and eventually lead to a creeping flow in a manner that is consistent with the regularized $\mu(I)$ rheology. Because the system will experience very low inertial numbers during deceleration and creeping flow, the regularization of the $\mu(I)$ rheology is crucial here to ensure numerical stability \cite{barker2017partial}.

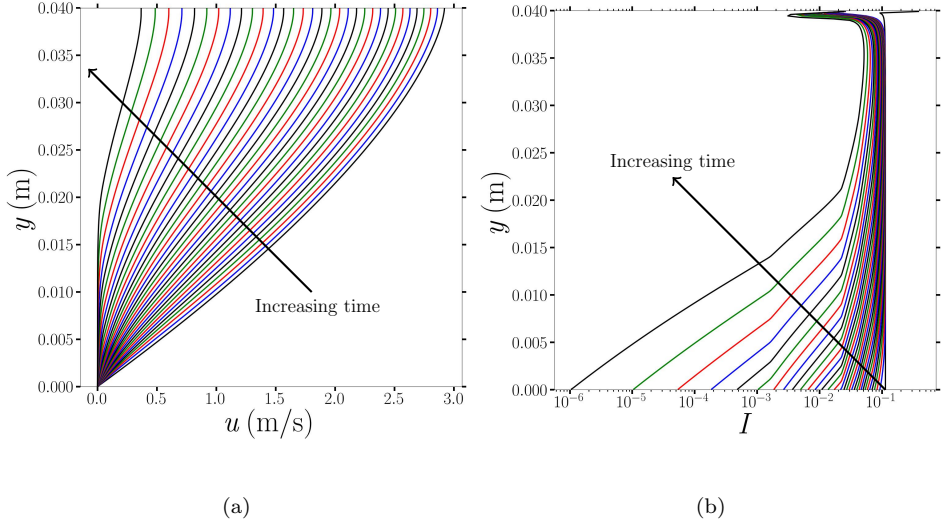
\begin{figure}[htbp]
    \centering
    \begin{subfigure}[b]{0.45\textwidth}
        \centering
        \includegraphics[width=\textwidth,keepaspectratio=true]{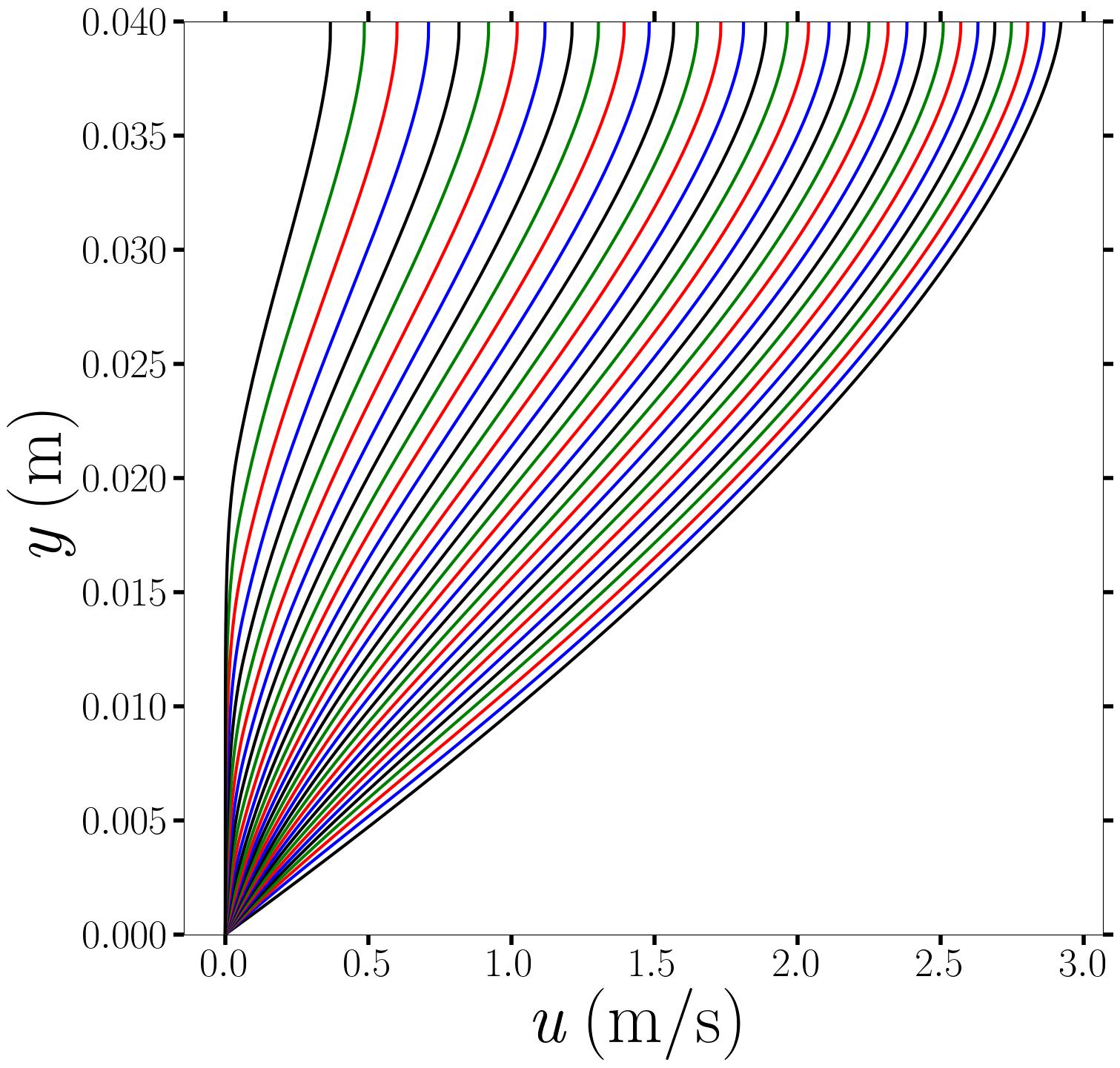}
        \begin{tikzpicture}[overlay, remember picture]
            \draw[->, black, thick] (1.0, 2.5) -- ++(135:4.2) node[anchor=south east,scale=0.6] at (2.0, 2.1) {Increasing time};
        \end{tikzpicture}
        \caption{}
    \end{subfigure}
    \begin{subfigure}[b]{0.45\textwidth}
        \centering
        \includegraphics[width=\textwidth,keepaspectratio=true]{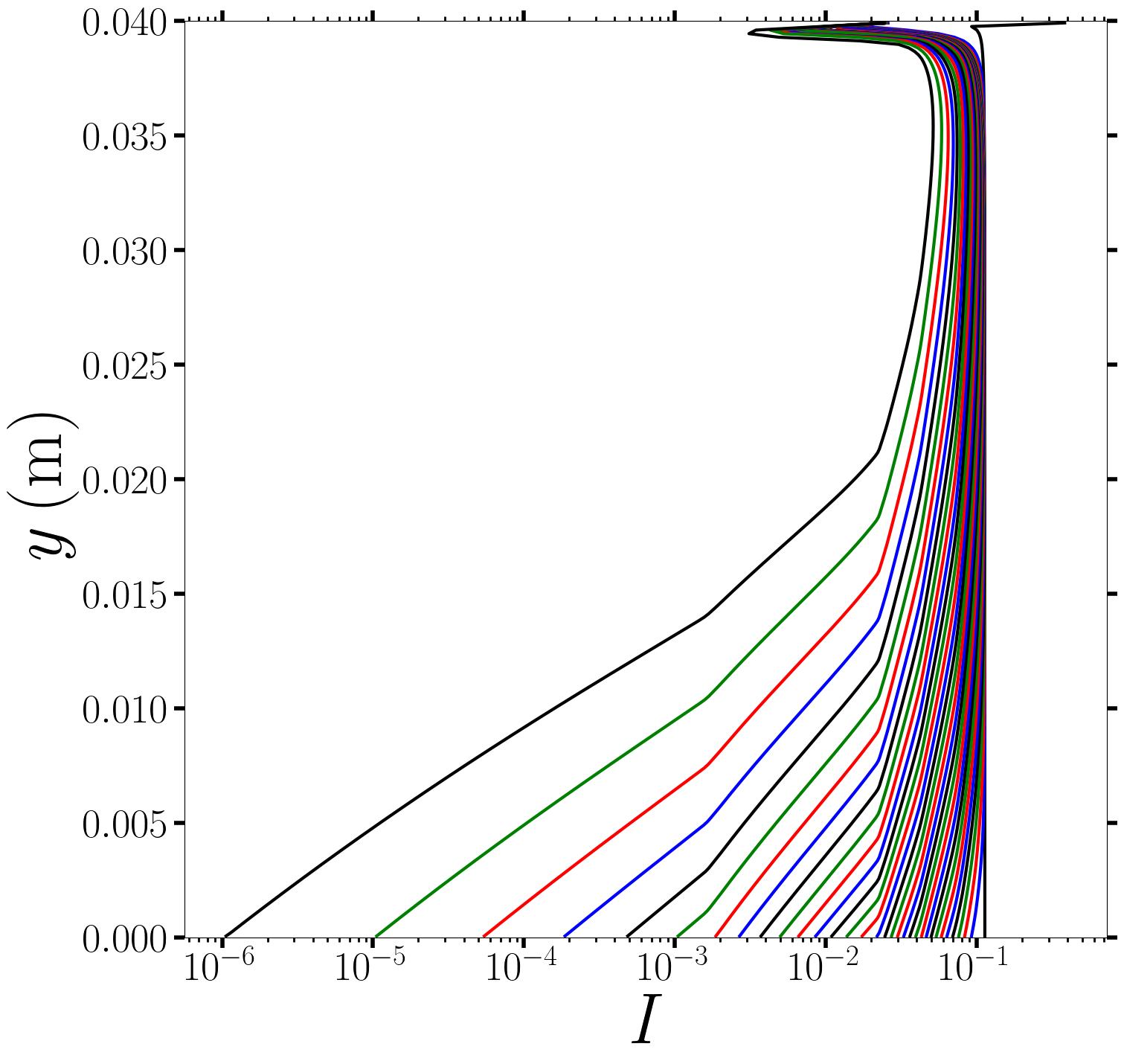}
        \begin{tikzpicture}[overlay, remember picture]
            \draw[->, black, thick] (2.32, 1.2) -- ++(135:4) node[anchor=south,scale=0.6] {Increasing time};
        \end{tikzpicture}
        \caption{}
    \end{subfigure}
    \caption{Decelerating chute flow initialized with a steady-state Bagnold's velocity profile at an inclination angle $\zeta_{0} = 0.43$, or $24.6^{\text{o}}$. The inclination is reduced to $\zeta = 0.175$, or $10^{\text{o}}$, at time $t=0$, which is considerably lower than the dynamic angle of repose $\text{arctan}(\mu_s)=18.3^{\text{o}}$ for this system. (a) Velocity $u$ and (b) inertial number $I$ profiles with respect to $y$ plotted at every $1000$ time steps up to $t\approx1$s.}
    \label{fig:double_layer_decelerate}
\end{figure}

Figure \ref{fig:double_layer_decelerate} shows the transient evolution of (a) velocity $u$ and (b) inertial number $I$ as a function of $y$, as the system responds to the reduced angle of inclination over a time of $t\approx 1$s corresponding to $3.2\times10^4$ steps. As time progresses, both $u$ and $I$ reduce considerably, and the region near the bottom wall becomes (nearly) static. The low cutoff inertial number in these simulations from the NN-ensemble is estimated to be $I^{N} \approx 1.6\times10^{-3}$; however, considerably lower inertial numbers $I\approx 10^{-6}$ are reached in the simulation as a result of regularization, thus demonstrating numerical stability of our multiscale modeling framework at very low inertial numbers. Similar to the previous example, the multiscale approach queried 12 new DEM simulations with $\kappa=5$ in addition to the 4 initial data points in $\bm{I}_{0}$, again demonstrating the computational efficiency of our adaptive hierarchical multiscale method.
We also ran the same numerical example with unregularized $\mu(I)$ rheology, and the simulation became unstable at a much earlier time $t\approx0.7$s for a much higher inertial number $I\approx5.3\times10^{-4}$.

\subsection{Sub-aerial granular column collapse}
\label{granular_column_collapse}
\begin{figure}[ht]
    \centering
    \includegraphics[width=0.95\textwidth,keepaspectratio=true,trim={4cm 7cm 4cm 7cm},clip]{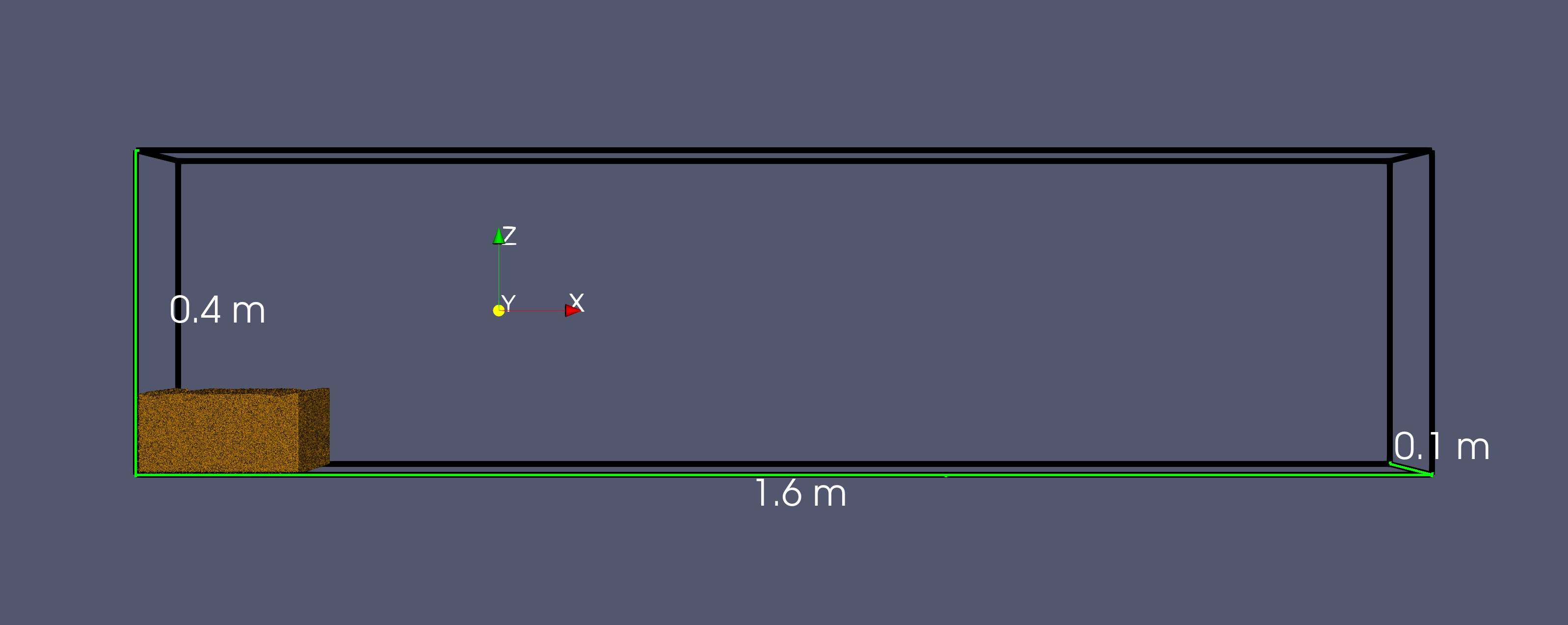}
    \caption{The setup for sub-aerial granular column collapse. A granular column of size $(L_{c}, W_{c}, H_{c})$ is initially held at rest. The dynamics of collapse are analyzed using continuum simulations for different column aspect ratios $H_{c}/L_{c}$ such that the total volume of the granular material $L_c\times W_c \times H_c = 2\times10^{-3}$m\textsuperscript{3} and $W_{c} = 0.1$m are fixed.}
    \label{fig:column_collapse}
\end{figure}

In the last numerical experiment, we consider a considerably more challenging setup of three-dimensional sub-aerial granular column collapse, which has been extensively studied previously using both DEM and continuum simulations. A granular column of length $L_{c}$, width $W_{c}$, and height $H_{c}$ in background air is initially held at rest in a domain of size $1.6$m$\times0.1$m$\times0.4$m, as shown in Fig. \ref{fig:column_collapse}. The density and viscosity of air are fixed at $\rho_f=1$kg/m$^{3}$ and $\eta_{f} = 1.48\times10^{-5}$Pa$\cdot$s respectively. No-slip boundary conditions are imposed at $z=0$, $z=0.4$m, and $x=1.6$m, whereas $x=0$ is treated as a full slip wall. The system is periodic in the $y$ direction and gravity acts in the negative $z$ direction. Here we vary the initial aspect ratio $H_c/L_c$ for the same volume of the granular column to study its impact on the column collapse dynamics, and in particular the column run-out distance.

In addition to a regularized $\mu(I)$ rheology that is adaptively learned from the NN-ensemble, in this example we also impose: (1) a minimum granular viscosity $\eta_g^{\text{min}}=1.6\times10^{-1}$Pa$\cdot$s, as described in previous numerical examples, and (2) a minimum value of strain rate $\epsilon_{|\bm{D}|}\approx 5 \times 10^{-3} \sqrt{g/H_{c}}$, such that the regularized granular viscosity is:
\begin{gather}
    \eta_g = \frac{\mu(I) \, p_{s}}{2 \left(|\bm{D}| + \epsilon_{|\bm{D}|}\right)}.
\end{gather}
While $\eta_{g}$ is expected to remain bounded and not diverge at very low inertial numbers due to regularization, in this numerical example we observed that without the inclusion of $\epsilon_{|\bm{D}|}$, the viscosity can become very large, which causes convergence issues for the linear solver used in the continuum simulations. While these issues can be resolved by using more sophisticated linear solvers, a creeping strain rate $\epsilon_{|\bm{D}|}$ has been commonly included in several previous continuum simulations of dense granular flows.

\subsubsection{Collapse dynamics and comparison with DEM simulations}
\begin{figure}[htbp]
    \centering
    \begin{subfigure}[b]{0.48\textwidth}
        \centering
        \includegraphics[width=\textwidth,keepaspectratio=true]{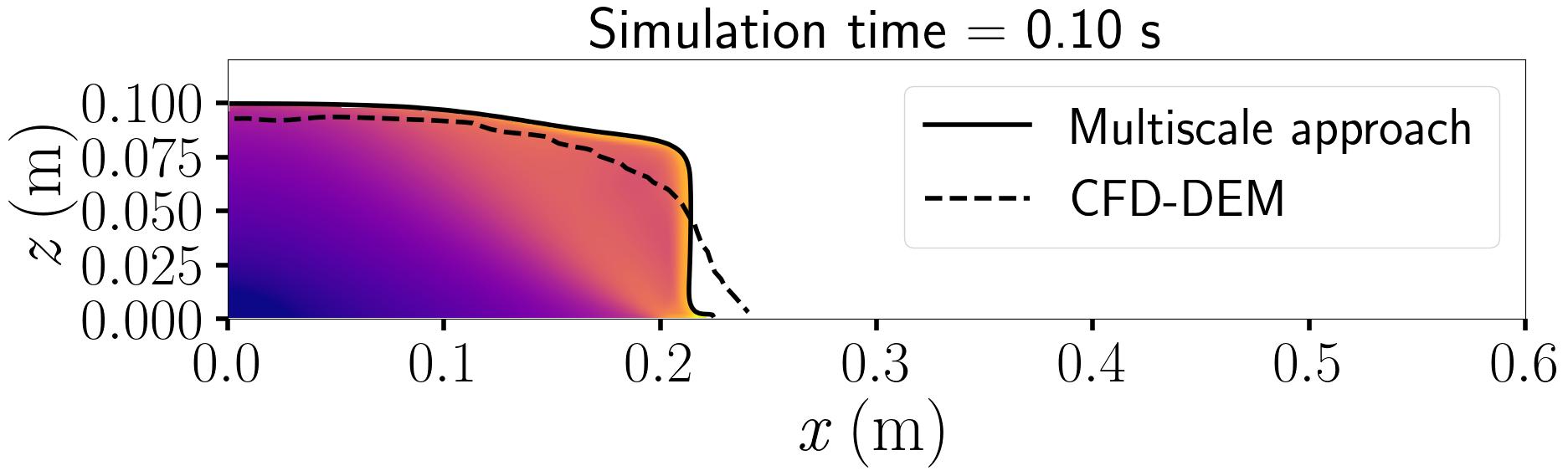}
    \end{subfigure}
    \begin{subfigure}[b]{0.48\textwidth}
        \centering
        \includegraphics[width=\textwidth,keepaspectratio=true]{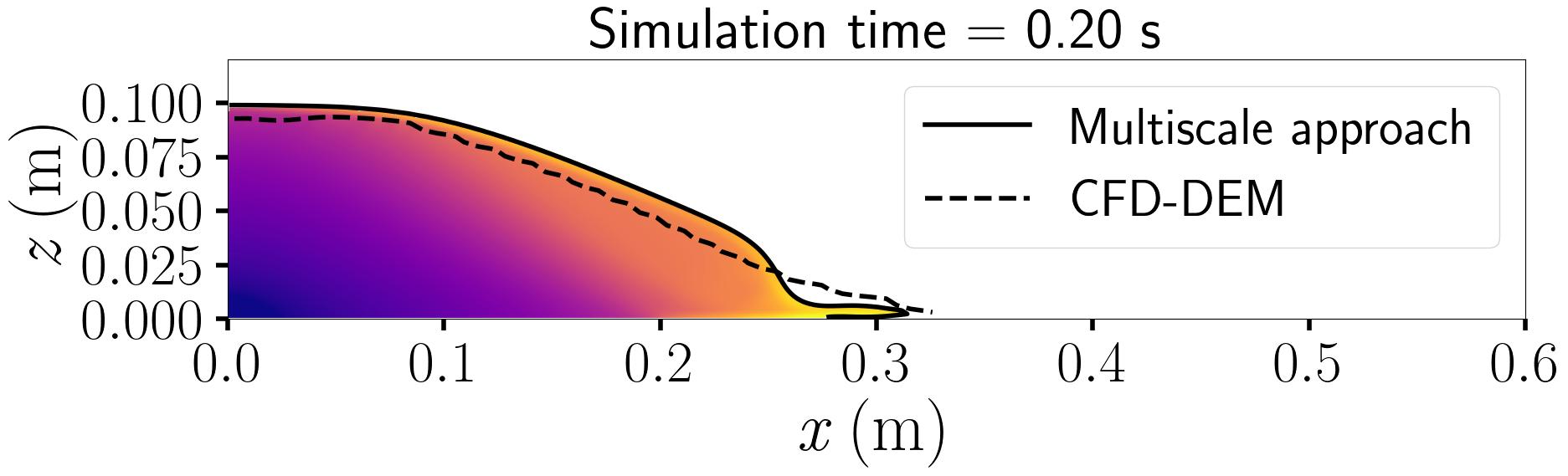}
    \end{subfigure}
    \begin{subfigure}[b]{0.48\textwidth}
        \centering
        \includegraphics[width=\textwidth,keepaspectratio=true]{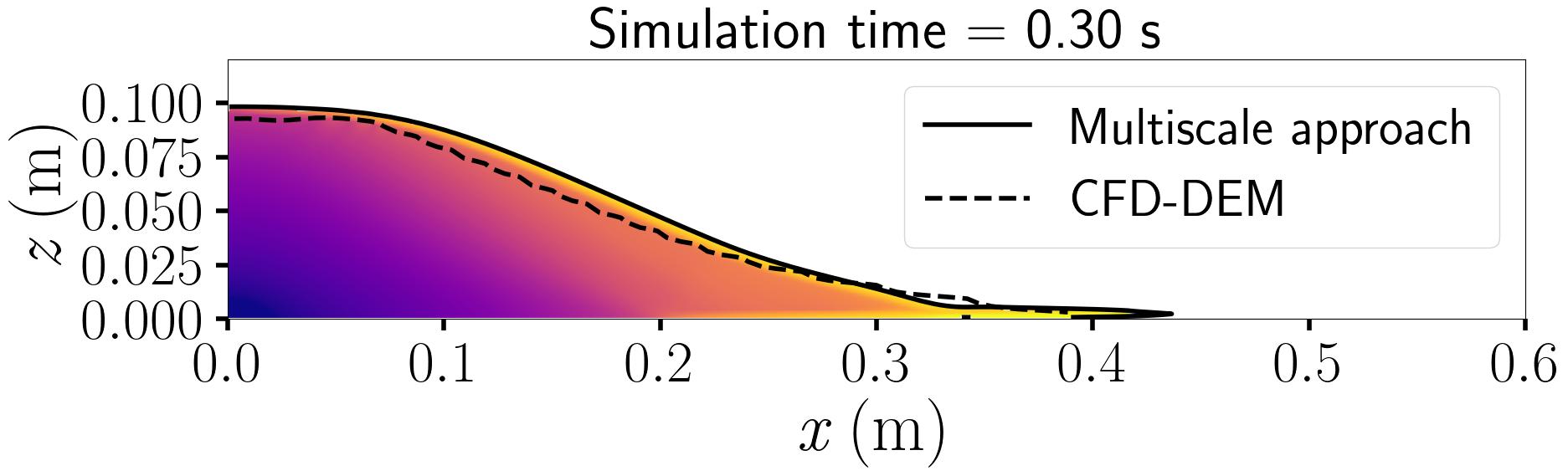}
    \end{subfigure}
    \begin{subfigure}[b]{0.48\textwidth}
        \centering
        \includegraphics[width=\textwidth,keepaspectratio=true]{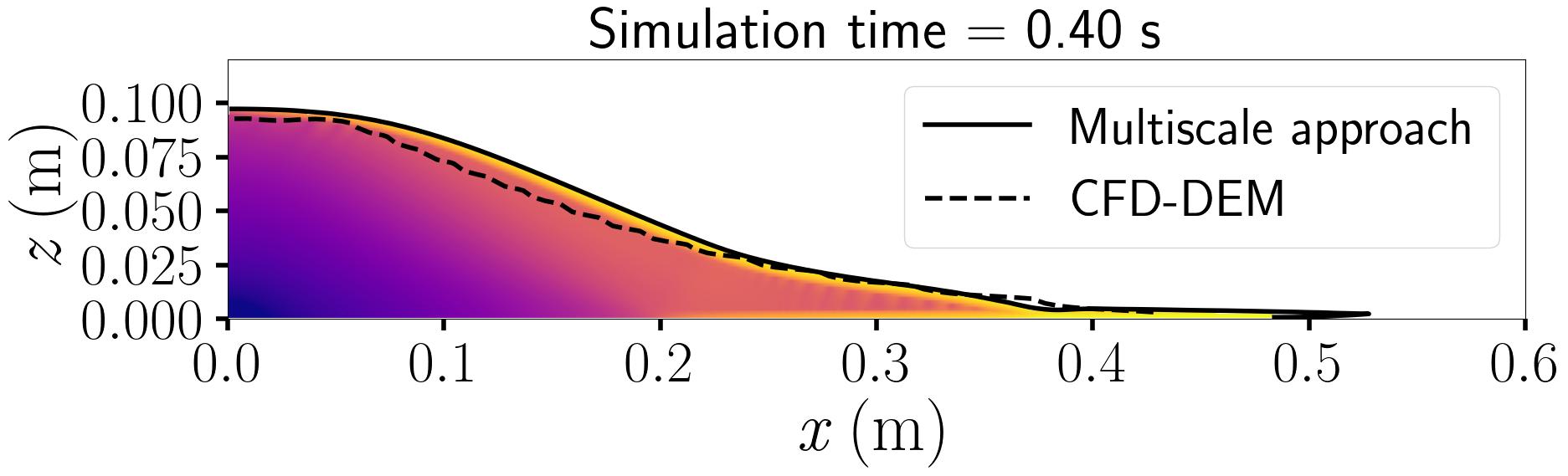}
    \end{subfigure}
    \centering
    \begin{subfigure}[b]{0.48\textwidth}
        \centering
        \includegraphics[width=\textwidth,keepaspectratio=true]{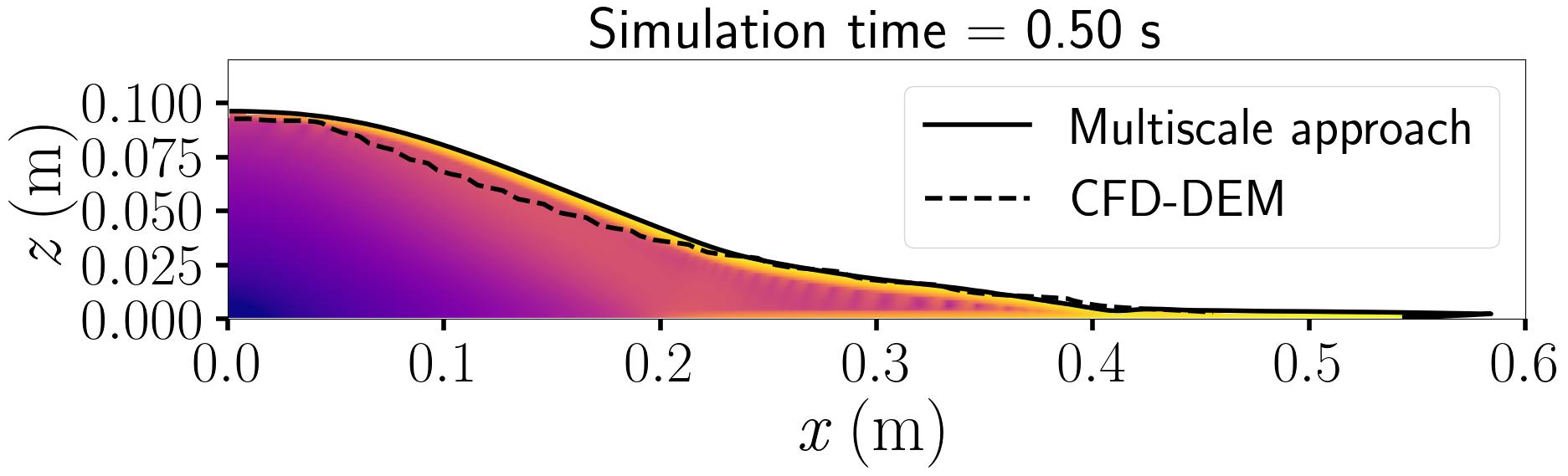}
    \end{subfigure}
    \begin{subfigure}[b]{0.48\textwidth}
        \centering
        \includegraphics[width=\textwidth,keepaspectratio=true]{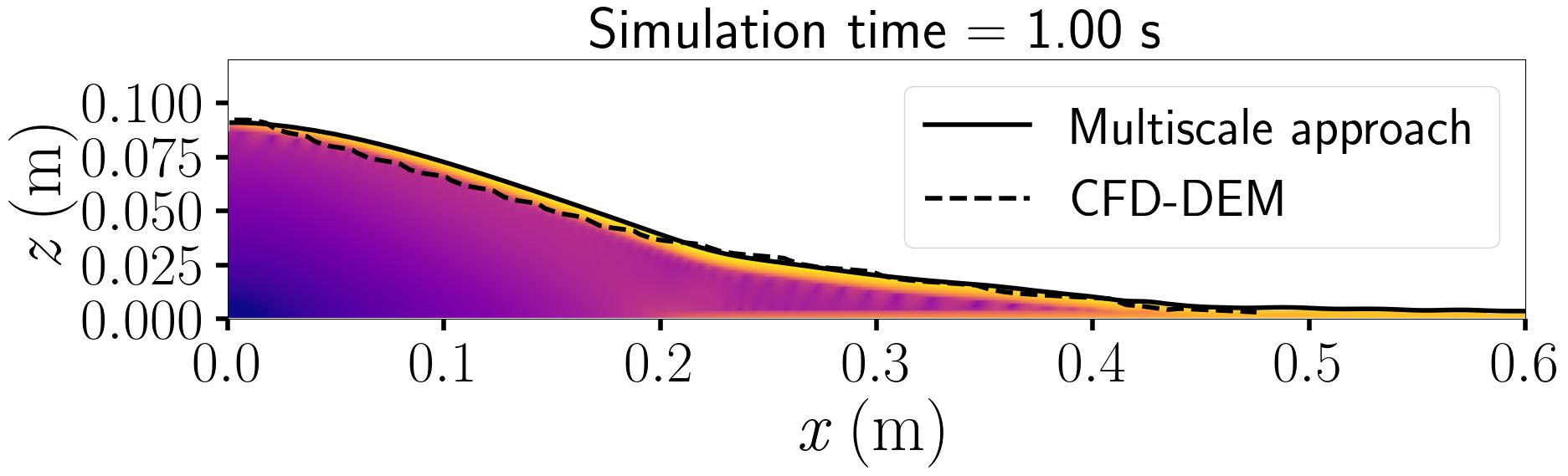}
    \end{subfigure}
    \begin{subfigure}[b]{0.5\textwidth}
        \centering
        \includegraphics[width=\textwidth,keepaspectratio=true]{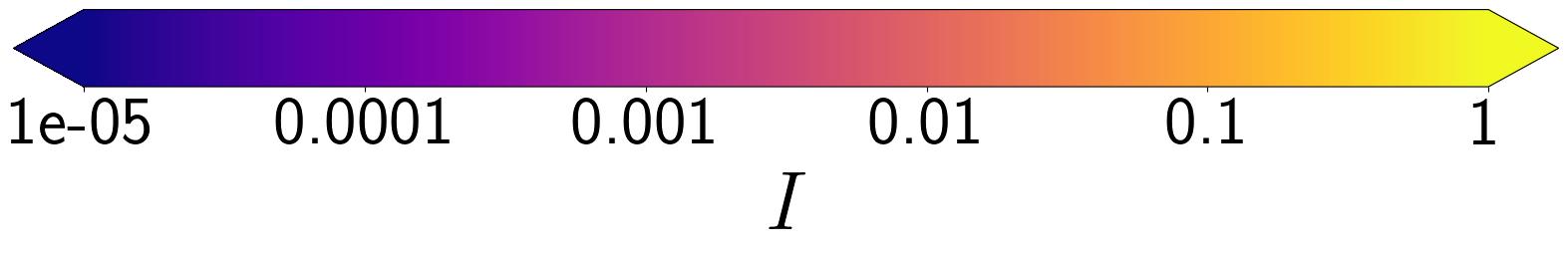}
    \end{subfigure}
    \caption{Comparison of granular column collapse dynamics between the adaptive hierarchical multiscale simulation and CFD-DEM simulation. The $c = 0.5$ contour line is shown for multiscale approach while $\phi_{p} = 0.5$ is shown for CFD-DEM. The figure also depicts local inertial number $I$ contour for $c \geq 0.5$ for the multiscale approach.}
    \label{fig:validation_contour}
\end{figure}

Figure~\ref{fig:validation_contour} shows the dynamics of the column collapse for $H_c/L_c=0.5$ at various simulation times. The interface between the air and granular material representing the surface of the granular column is depicted using a contour line corresponding to $c=0.5$. The granular material below the free surface for $c\geq0.5$ is colored by the local inertial number. The system is discretized on a uniform $1024\times64\times256$ grid, and the simulation is run with a CFL of 0.35 for a total simulation time of $t=1$s. In \ref{appendix_grid_sensitivity}, we discuss the sensitivity of the column collapse to the grid size.

At early times, the collapse of the column is rapid, as also seen by high inertial numbers in the rapidly deforming regions of the granular material. At later times, the collapse of the column is significantly slower, wherein most of the granular column is nearly static, i.e., creeping at very low inertial numbers. In the diffuse interface formulation of the two-component mixture considered here, a thin region of high inertial numbers is observed at the free surface due to high shear rate and low hydrostatic pressure. This issue can be ameliorated by using interface-tracking methods such as the volume-of-fluid approach described in Rauter \cite{rauter2021compressible}. 

We compare the dynamics of column collapse from our multiscale simulation framework with an Euler-Lagrange simulation where the air is treated as a continuum and particles are evolved as discrete entities. The simulation is conducted using an open-source computational fluid dynamics-discrete element method (CFD-DEM) solver, MFIX-Exa \cite{musser2022mfix,fullmer2025dense}. The numerical details and the setup of the CFD-DEM simulation are described in  \ref{appendix_cfd_dem}. The free surface of the granular column is tracked by a contour line corresponding to the local particle volume fraction $\phi_p=0.5$. At early times $t = $ 0.1s and 0.2s, we observe a noticeable difference between the fronts of the column collapse computed with CFD-DEM and multiscale modeling, which are reduced at later times. The differences at early times can be attributed to the differences in the treatment of the $x=0$ boundary. The multiscale model treats this boundary as no-slip while the CFD-DEM simulation includes a Coulomb slip condition in particle-wall interactions \cite{fullmer2025dense}. A frictional boundary instead of a no-slip boundary could potentially provide a better representation of the granular dynamics near the wall \cite{rauter2021compressible}.

\subsubsection{Collapse run-out for different column aspect ratios}
Next, we analyze the dependence of column collapse run-out distance as a function of initial granular column aspect ratio. For this purpose, we used the NN-ensemble trained on $H_{c}/L_{c} = 0.5$ simulation discussed above as a pre-trained model for running simulations of different initial aspect ratios. We remark that the ability to re-use a trained NN-ensemble on a different problem is a significant benefit of the adaptive multiscale modeling framework. In the present case, the model was sufficiently well-trained in the original multiscale simulation that no new DEM simulations were needed in any of the new simulations with different aspect ratios.

\begin{figure}[htbp]
    \centering
    \includegraphics[width=0.55\linewidth,keepaspectratio=true]{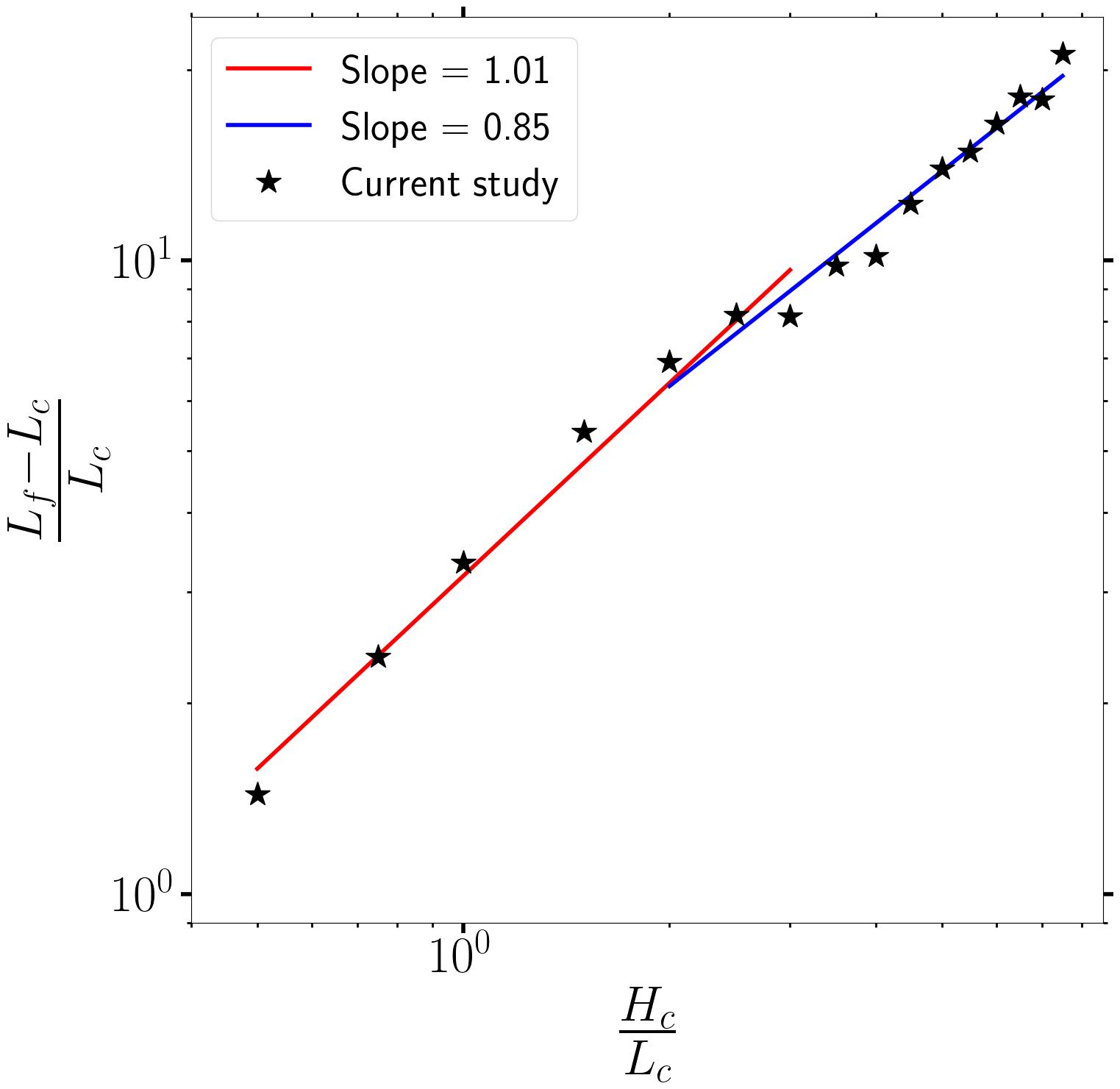}
    \caption{The normalized run-out distance, $(L_f-L_c)/L_c$, in granular column collapse as a function of the initial column aspect ratio, $H_c/L_c$. The run-out distance scales nearly linearly with $H_c/L_c$ at low aspect ratios, and it transitions to a power-law scaling with an exponent of $0.85$ at $H_c/L_c\approx 2$ for high aspect ratios.
    }
    \label{fig:runout}
\end{figure}

Figure \ref{fig:runout} shows the variation of run-out distance $L_f$, normalized by the initial length of the column defined by $(L_f-L_c)/L_c$, as a function of the initial column aspect ratio $H_c/L_c$. All the simulations were run until a total simulation time of $5 \sqrt{H_c/g}$, and we confirmed that after this time all the collapsed columns were (nearly) static in most regions of the granular material. The run-out distance $L_f$ was computed as the $x$-intercept of a linear fit for the $c=0.5$ contour line. The thin, elongated granular material region close to the $z=0$ wall does not represent the characteristics of the bulk granular material. Our methodology for estimating run-out distance is relatively insensitive to the behavior in this region. At low aspect ratios, $(L_f-L_c)/L_c$ scales nearly linearly with $H_c/L_c$, which is consistent with previous studies \cite{staron2005study,dunatunga2015continuum,chantharayukhonthorn2025hybrid}. At high aspect ratios transitioning around $H_c/L_c \approx 2$, we observe a power-law scaling with an exponent of $0.85$, which is slightly higher than previous reported values \cite{staron2005study,dunatunga2015continuum}. We postulate that incorporating interface-tracking methods such as the volume-of-fluid will facilitate a more accurate identification of the run-out distance.

\subsubsection{Adaptive evolution of the neural network ensemble}
Using the column collapse simulation with an initial aspect ratio of $0.5$ as a representative example, we now describe how the NN-ensemble model with $\kappa=5$ adaptively evolves during the course of a continuum simulation. We also demonstrate the computational efficiency of our approach in accurately capturing the $\mu(I)$ rheology using minimal DEM simulations.

\begin{figure}[htbp]
    \centering
    \begin{subfigure}[b]{0.32\textwidth}
        \centering
        \includegraphics[width=\textwidth,keepaspectratio=true]{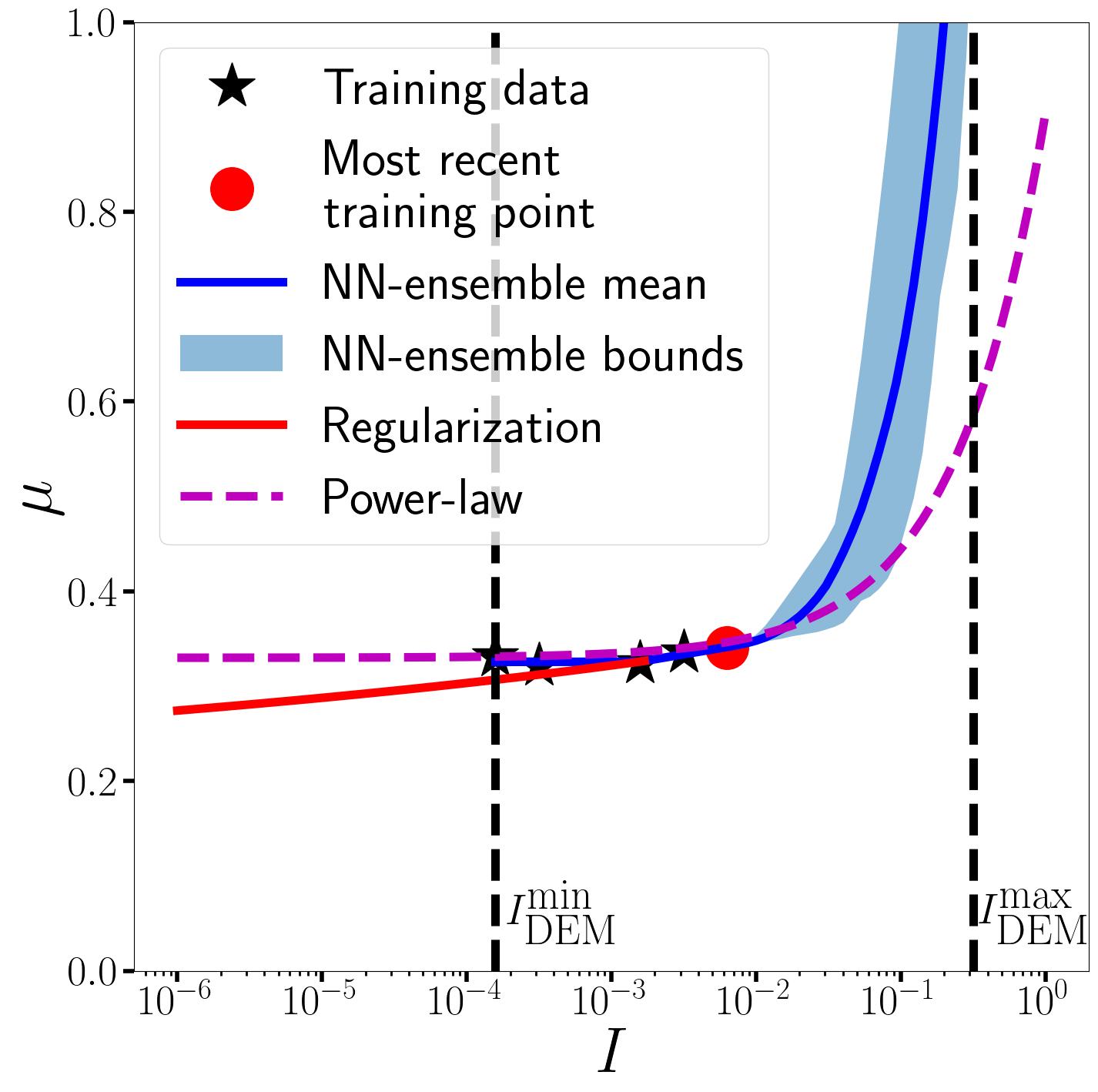}
        \caption{Continuum step number = 1.}
    \end{subfigure}
    \begin{subfigure}[b]{0.32\textwidth}
        \centering
        \includegraphics[width=\textwidth,keepaspectratio=true]{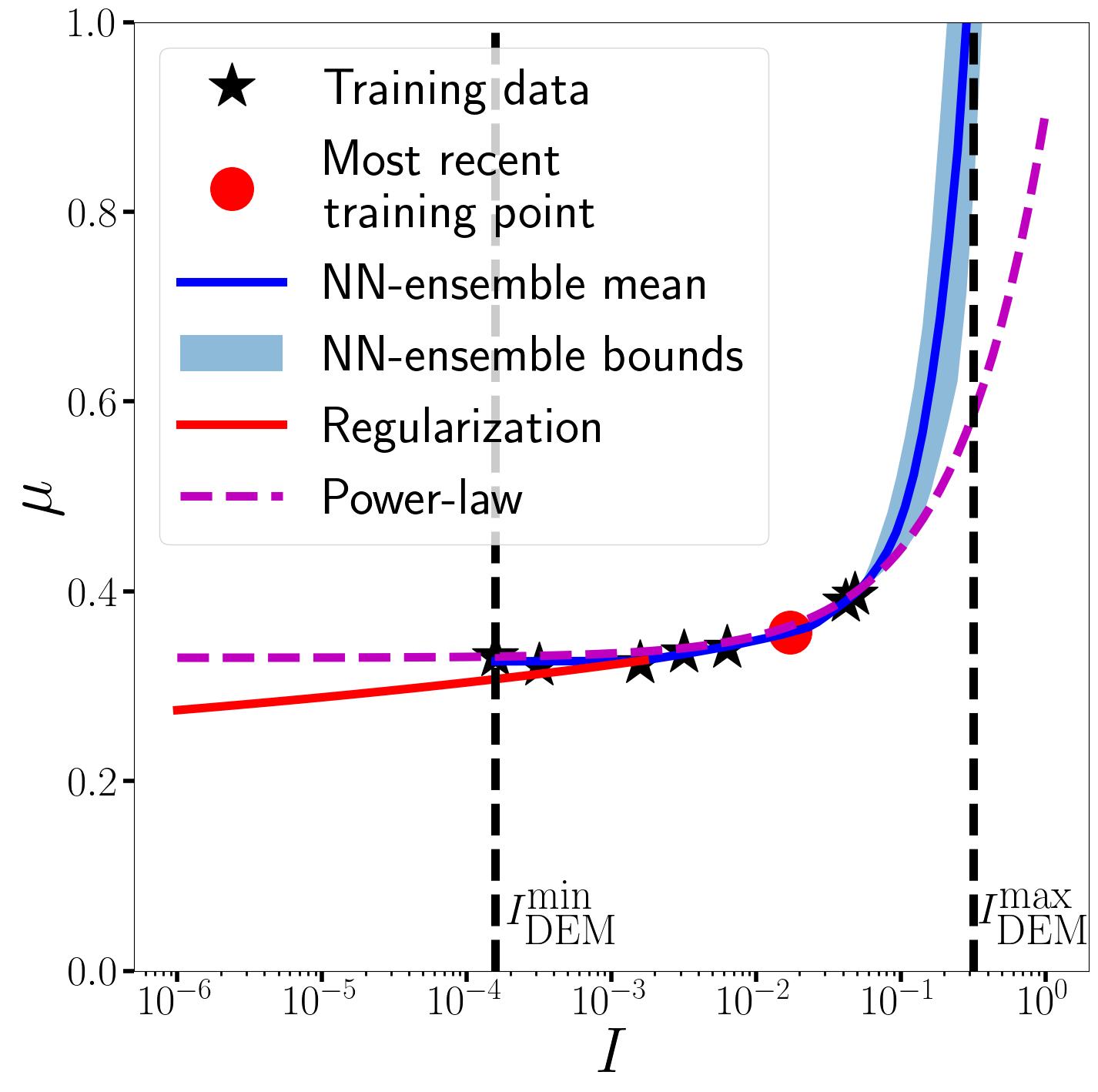}
        \caption{Continuum step number = 2.}
    \end{subfigure}
    \begin{subfigure}[b]{0.32\textwidth}
        \centering
        \includegraphics[width=\textwidth,keepaspectratio=true]{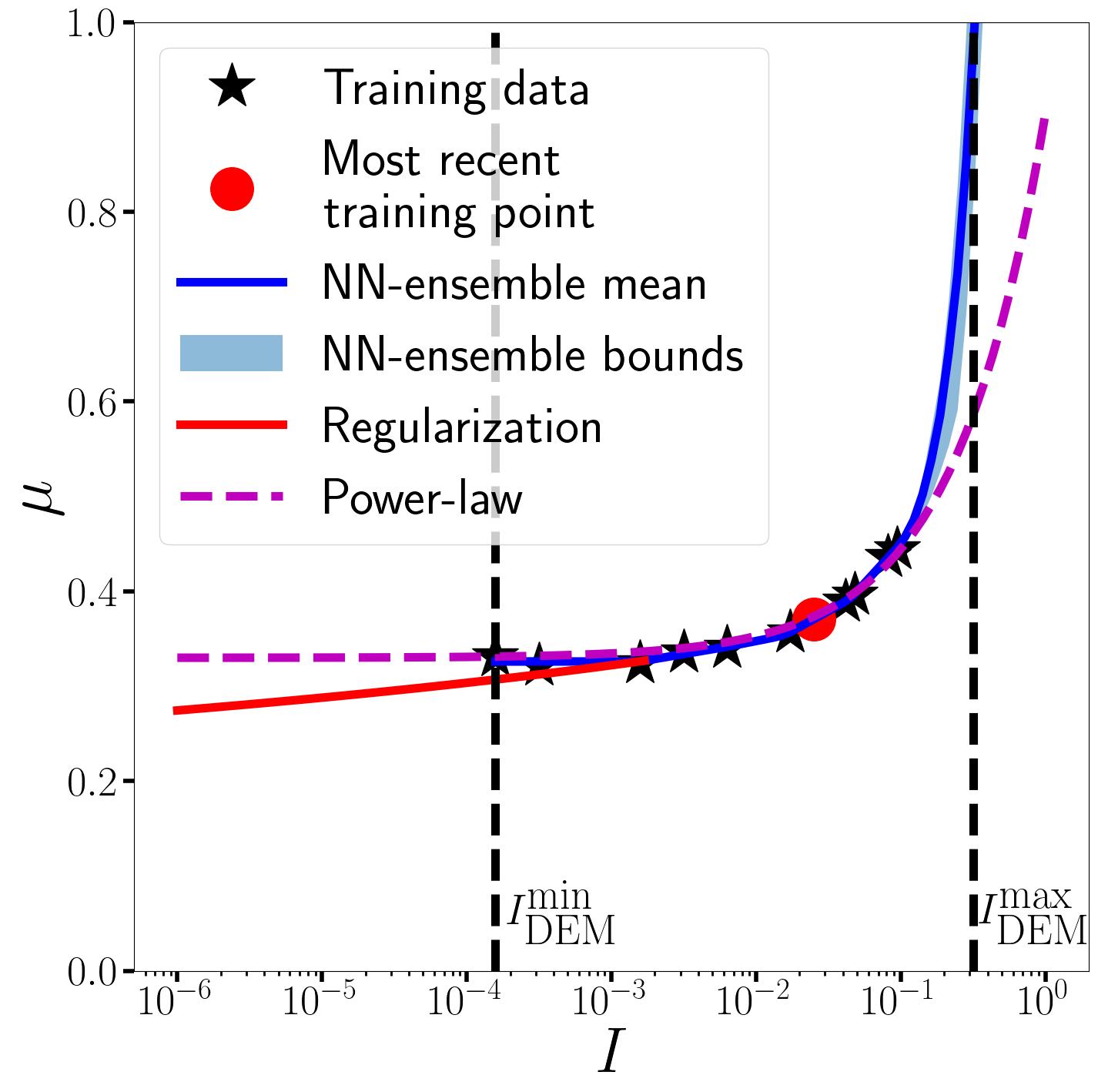}
        \caption{Continuum step number = 3.}
    \end{subfigure}
    \begin{subfigure}[b]{0.32\textwidth}
        \centering
        \includegraphics[width=\textwidth,keepaspectratio=true]{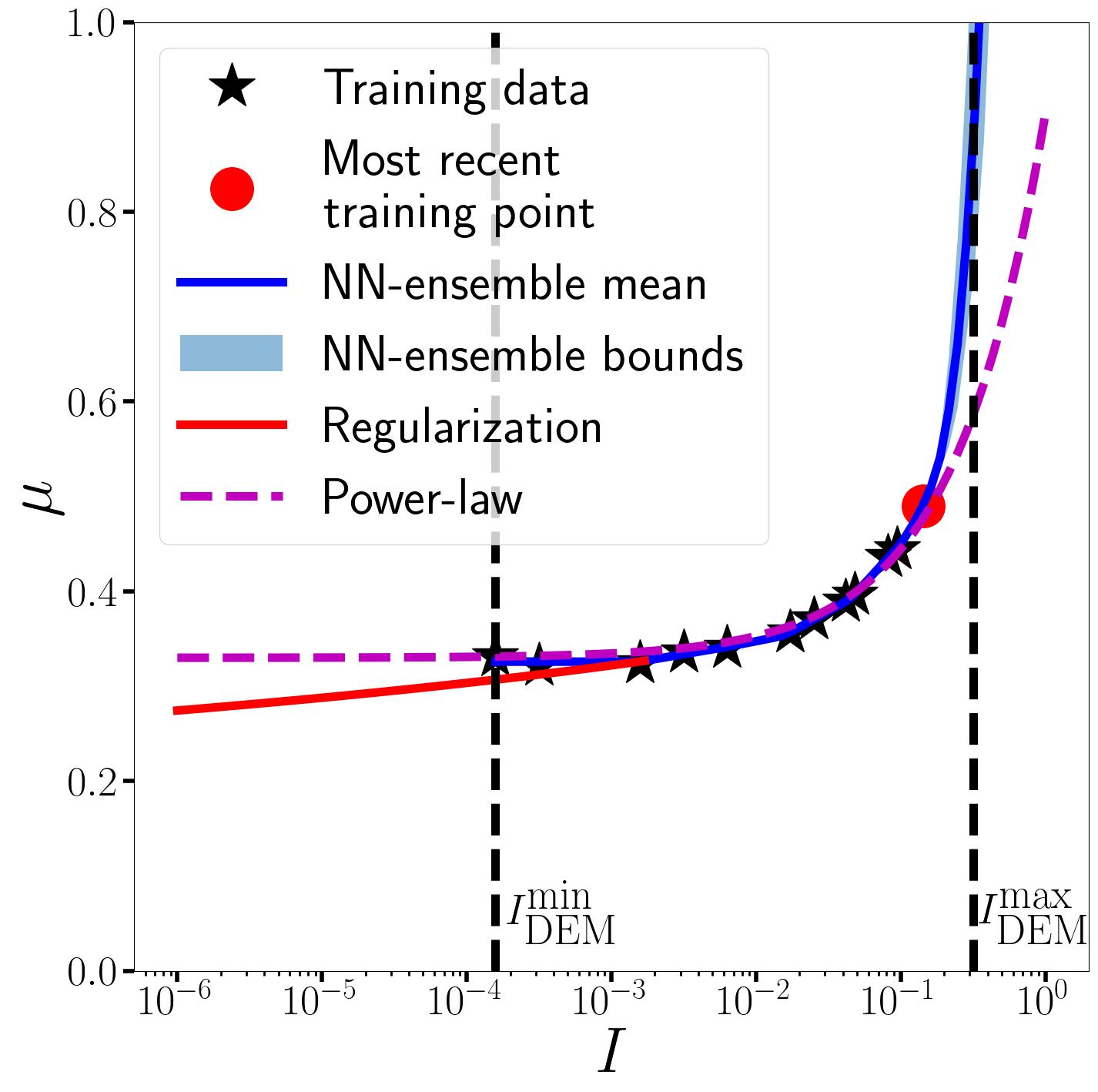}
        \caption{Continuum step number = 4.}
    \end{subfigure}
    \centering
    \begin{subfigure}[b]{0.32\textwidth}
        \centering
        \includegraphics[width=\textwidth,keepaspectratio=true]{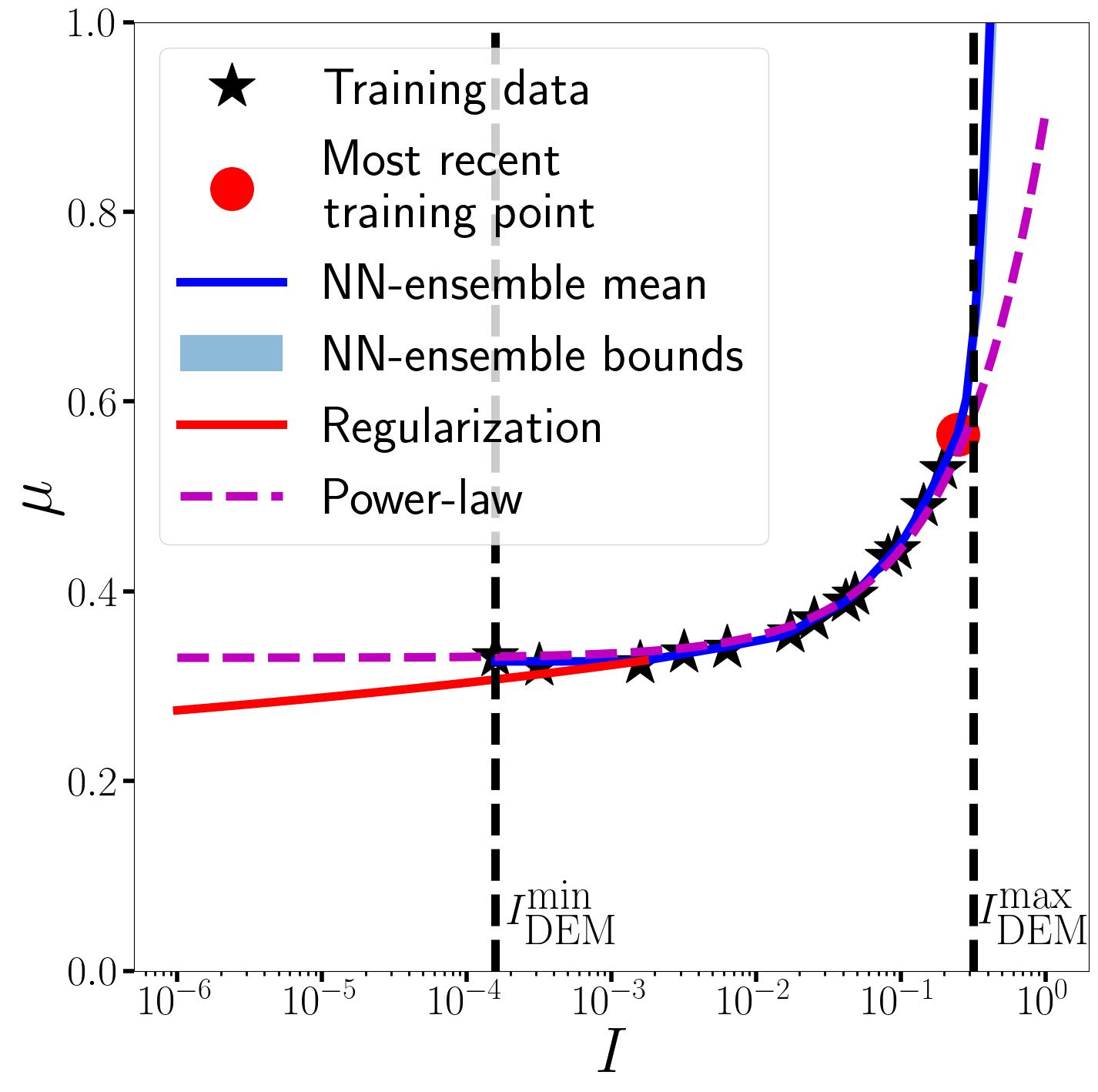}
        \caption{Continuum step number = 6.}
    \end{subfigure}
    \begin{subfigure}[b]{0.32\textwidth}
        \centering
        \includegraphics[width=\textwidth,keepaspectratio=true]{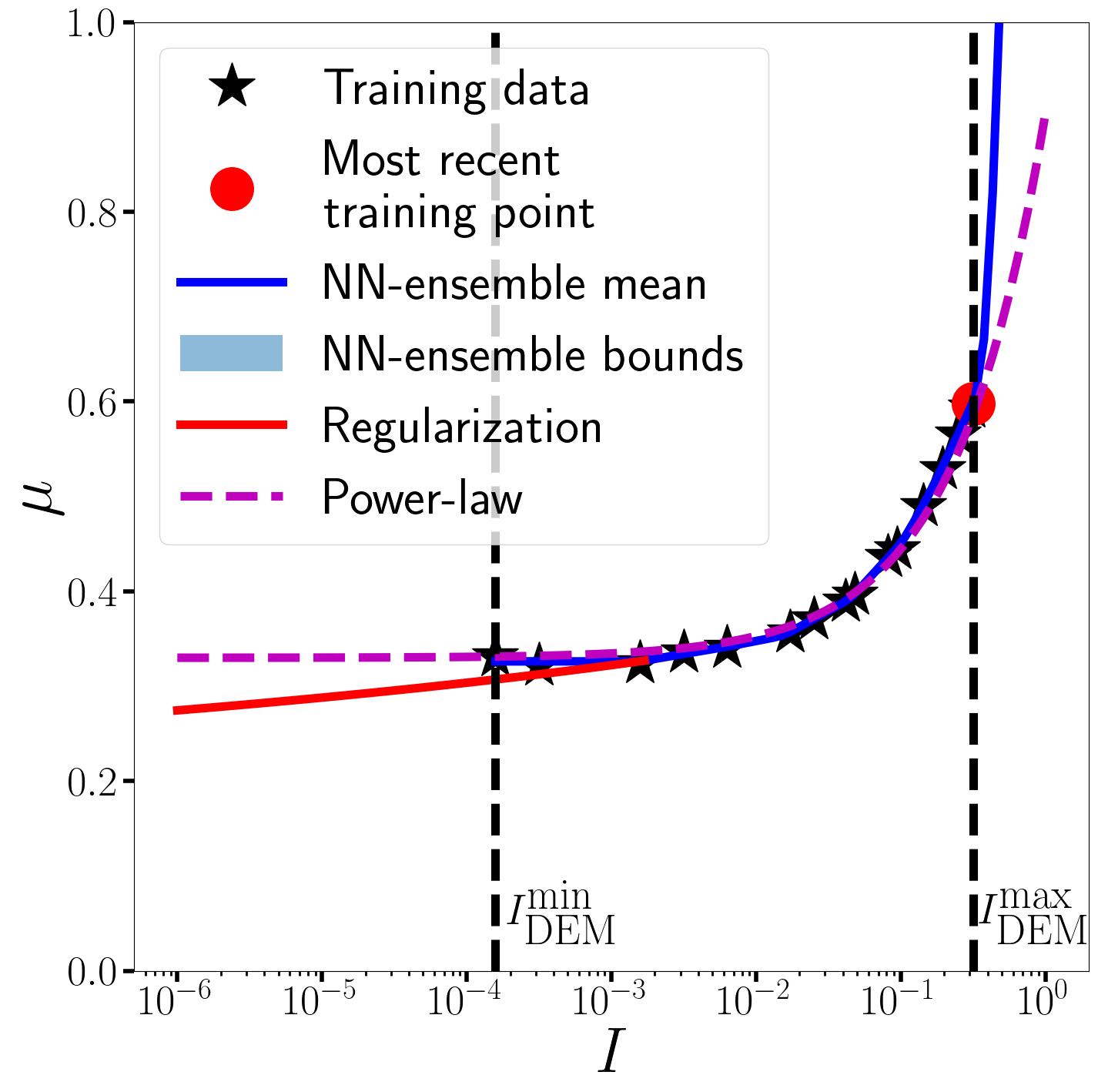}
        \caption{Continuum step number = 8.}
    \end{subfigure}
    \caption{Evolution of the $\mu(I)$ rheology predicted by NN-ensemble model during an adaptive multiscale simulation of granular column collapse. (a)-(f) chronologically correspond to the continuum time steps at the initial stage of the simulation. The training dataset until the indicated time step is plotted with black asterisks, and the last training inertial number is shown with a red circle. The blue solid line corresponds to the mean $\mu(I)$ rheology predicted from the NN-ensemble model, and the blue shaded area bounds the minimum and maximum values of the model prediction. The black vertical dashed lines indicate the range within which DEM simulations are run. The red solid line corresponds to the regularized rheology that replaces the model prediction at low inertial numbers. The magenta dashed line shows the $\mu(I)$ rheology obtained from literature for a similar granular system \cite{srivastava2021viscometric}.}
    \label{fig:model_evolution}
\end{figure}

During the initial time steps of the column collapse, the simulation encounters relatively higher inertial numbers that were not observed previously. Therefore, DEM simulations are queried during these steps to update the NN-ensemble model and eventually provide the stress ratio for these inertial numbers. Figure \ref{fig:model_evolution} shows the evolution of the mean $\sum_{m=1}^{M}\mu_{m}/M$, and lower $\min(\{\mu_{m}\}_{m=1}^{M})$ and upper $\max(\{\mu_{m}\}_{m=1}^{M})$ of the NN-ensemble model at initial continuum time steps. Each panel of Fig. \ref{fig:model_evolution} shows the training data from DEM simulations that is used to train the NN-ensemble at each time step (black asterisks), where the last data point of the training set is specifically depicted with a red circle. The number of DEM simulations needed at each time step is variable; in the present example, time steps $2$ and $3$ each required $3$ new DEM simulations, whereas steps 1 and $4-8$ required only a single DEM simulation. It is apparent that as the multiscale simulation proceeds, the NN-ensemble model provides increasingly better prediction of the rheology, as seen by the decrease in the uncertainty bounds of the model. The rheology predicted by the NN-ensemble model is also in excellent agreement with the literature values for a similar system composed of frictional, monodisperse spheres \cite{srivastava2021viscometric,clemmer2021shear}, as shown in Fig. \ref{fig:model_evolution}. The slight discrepancy at very low inertial numbers can be attributed to the large stress fluctuations in DEM simulations at these inertial numbers. Commensurately, the regularization of the model at low inertial numbers also evolves with the simulation where the cutoff inertial number $I^{N}$ is also updated when the NN-ensemble is re-trained. Furthermore, at all times during the evolution of the NN-ensemble model, the desired properties of monotonicity and convexity are globally preserved. In this example, after time step $8$, no new DEM simulations are needed as the NN-ensemble model adequately predicts the rheology for $\kappa = 5$ throughout the rest of the continuum simulation.

\begin{figure}[htbp]
    \centering
    \begin{subfigure}[b]{0.32\textwidth}
        \centering
        \includegraphics[width=\textwidth,keepaspectratio=true]{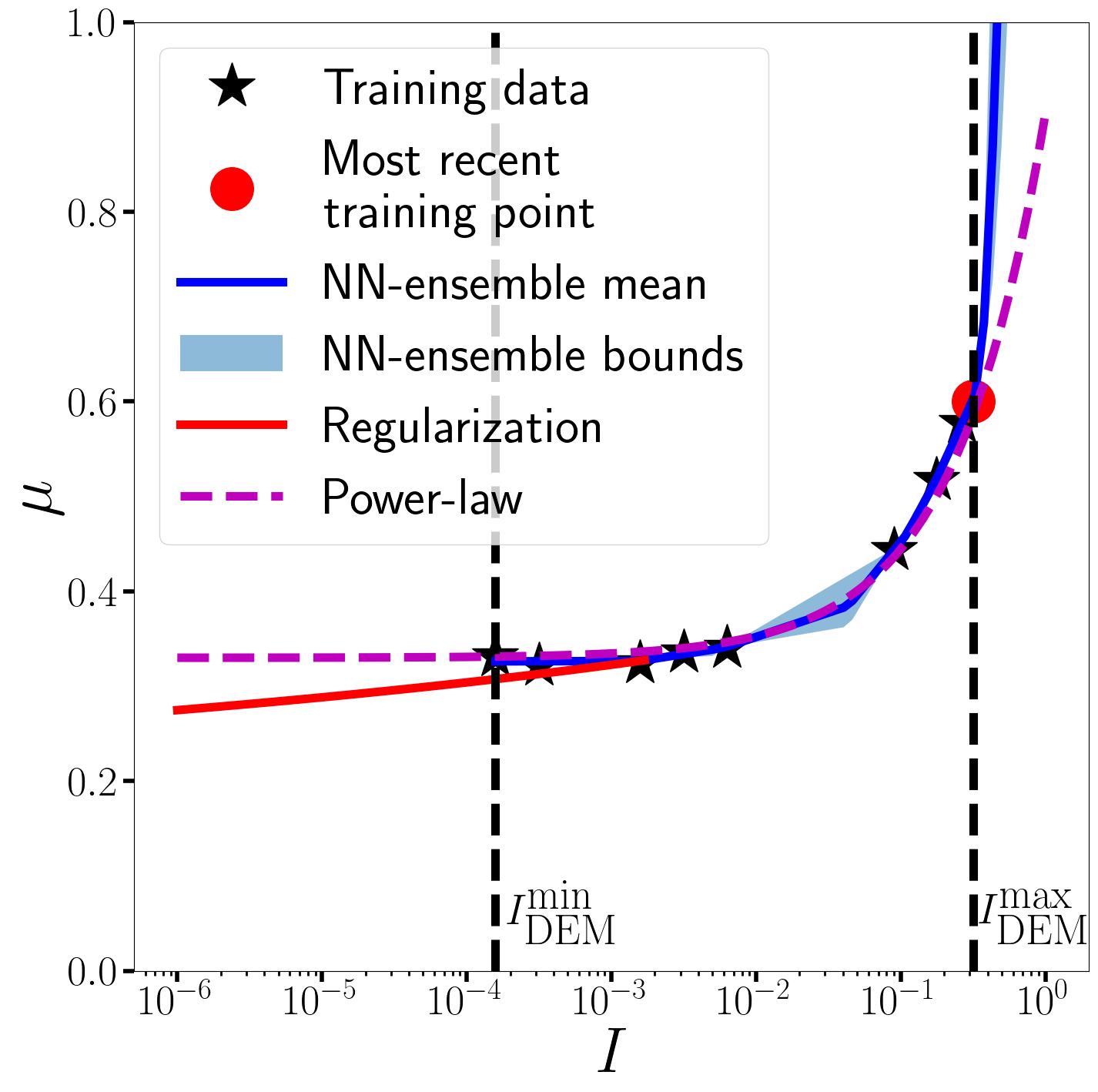}
        \caption{$\kappa = 100$.}
    \end{subfigure}
     \begin{subfigure}[b]{0.32\textwidth}
        \centering
        \includegraphics[width=\textwidth,keepaspectratio=true]{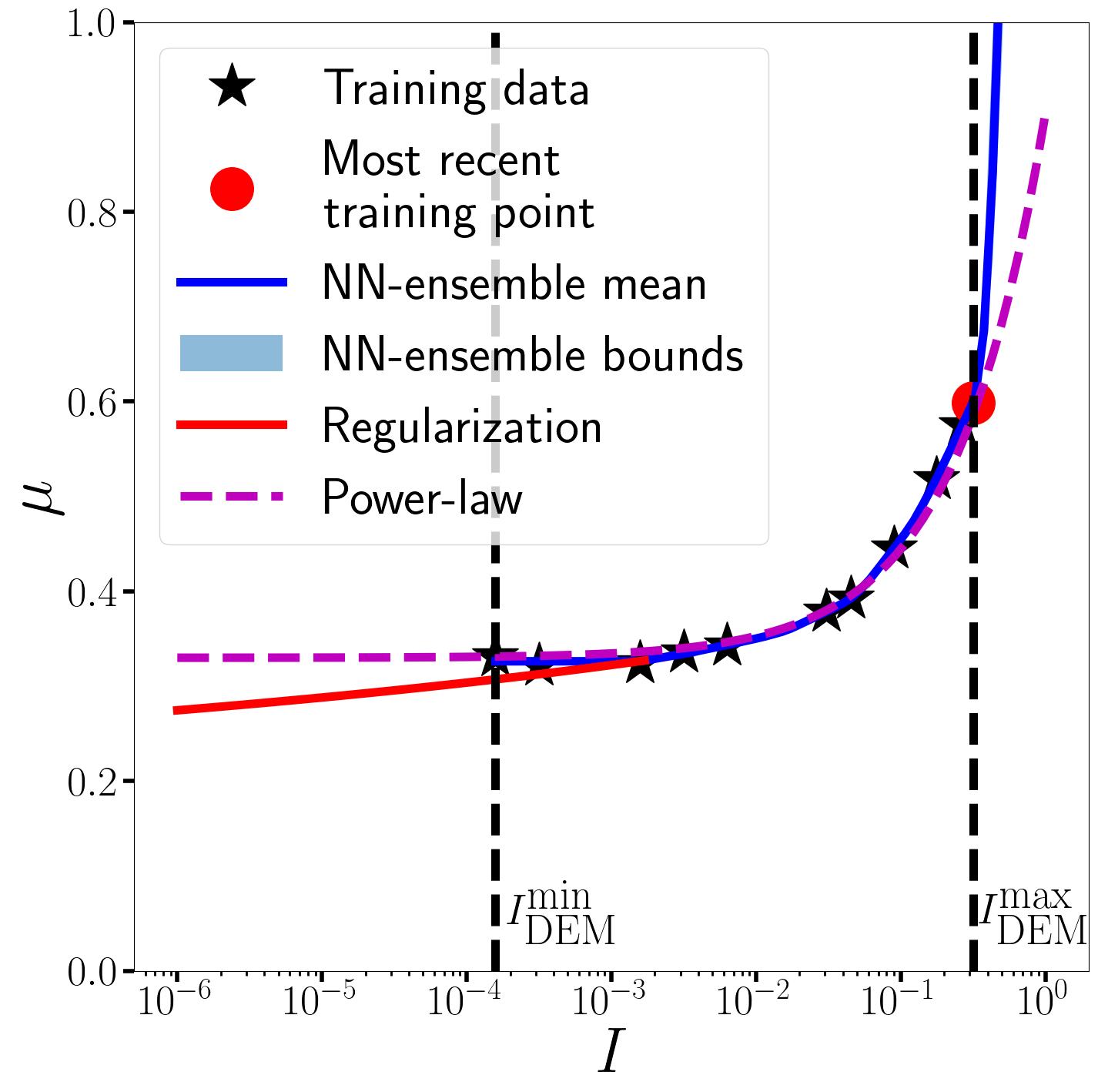}
        \caption{$\kappa = 20$.}
    \end{subfigure}
    \begin{subfigure}[b]{0.32\textwidth}
        \centering
        \includegraphics[width=\textwidth,keepaspectratio=true]{figures/fctr_5.0_a_0.5_grid_016_013.jpg}
        \caption{$\kappa = 5$.}
    \end{subfigure}
    \begin{subfigure}[b]{0.75\textwidth}
        \centering
        \includegraphics[width=\textwidth,keepaspectratio=true]{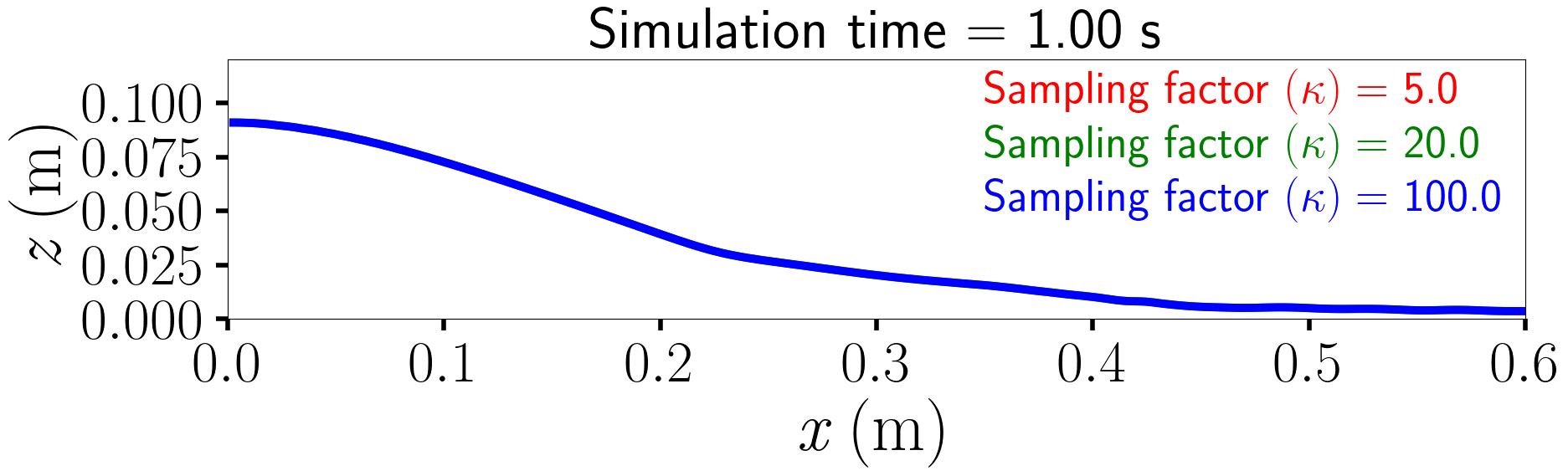}
        \caption{}
    \end{subfigure}
    \caption{(a) - (c) The final trained NN-ensemble model for $\mu(I)$ rheology for three different values of $\kappa$ obtained during adaptive hierarchical multiscale simulations of granular column collapse with the same initial column aspect ratio $H_c/L_c=0.5$. The details of various markers and plot lines are the same as described in Fig. \ref{fig:model_evolution}. (d) The $c\!=\!0.5$ contour lines depicting granular column collapse at the simulation end time $t=1.0$s for the three values of $\kappa$. The three lines are overlapping indicating minimal sensitivity of the collapse dynamics to $\kappa$.}
    \label{fig:sampling_sensitivity}
\end{figure}

We further investigate how the sampling factor $\kappa$ controls the evolution of NN-ensemble as well as column collapse dynamics by considering two additional multiscale simulations with $\kappa = 20$ and $\kappa=100$ conducted on the same simulation setup. An increase in $\kappa$ results in a lower number of DEM simulations that are required to train NN-ensemble model; a total of $12$ DEM simulations for $\kappa=5$, $7$ DEM simulations for $\kappa=20$, and $5$ DEM simulations for $\kappa=100$ were needed in this example. Fig. \ref{fig:sampling_sensitivity}(a) - (c) shows the fully-evolved NN-ensemble model for the three values of $\kappa$. While the mean value of the $\mu(I)$ rheology is similar across the three cases, $\kappa=100$ has a higher uncertainty bounds, as expected, resulting from a lower number of DEM simulations during the online training. However, the impact of $\kappa$ on the collapse dynamics appears to be negligible. Figure \ref{fig:sampling_sensitivity}(d) shows that the contours of column collapse at the same simulation end time $t=1$s for the three cases are nearly overlapping. While this result implies that $\kappa=100$ was a suitable choice in this example that would have utilized fewer DEM simulations, we postulate that more complex dense granular flows in complex geometries may require a stricter tolerance of uncertainty in the prediction of the rheology by NN-ensemble model. The study of such complex granular flows using our adaptive hierarchical multiscale method will be the subject of a future study.

\section{Conclusions}
\label{sec:conclusions}
This work presents an adaptive hierarchical multiscale method that couples a coarse-grained continuum incompressible flow solver with a fine-grained DEM solver to simulate dry and sub-aerial dense granular flows at practical length scales. The coupling itself is achieved by using an ensemble of neural networks that adaptively train for dense granular rheology using the DEM solver during a continuum simulation in a manner that is highly computationally efficient and also preserves the high-fidelity representation of the granular dynamics. Our customized neural network is structured to impose desired rheological properties such as monotonicity and convexity during training, in addition to incorporating regularization at low inertial numbers for numerical stability. We demonstrated the accuracy and efficiency of our multiscale method by first validating against known theory in $2$D steady-state inclined chute flows, and then tackling the more challenging cases of decelerating chute flows and $3$D sub-aerial granular column collapse.

The multiscale method described in this work is developed to adaptively learn the scalar $\mu(I)$ rheology, and as such is tailored to model quasi-steady, shear-dominated, dense granular flows that are well-described by such a rheology. However, our adaptive, data-driven approach utilizing neural networks is easily extensible to more complex granular flows that require an expanded set of input states and output rheological variables for their representation, such as anisotropic granular rheology with normal stress effects \cite{srivastava2021viscometric,sun2011constitutive} and Lode-angle dependent granular rheology \cite{clemmer2021shear}. Our method can also be extended to simulate transient granular flows where the rheology requires modeling the evolution of an additional internal state variable such as the fabric tensor \cite{sun2011constitutive,goddard2006dissipative,rojas2019capturing}.

While we utilized an NN-ensemble model for quantifying the uncertainty in the prediction of the dense granular rheology, the computational efficiency of our adaptive hierarchical multiscale model may be further enhanced by considering alternative probabilistic approaches such as the Monte Carlo dropout method \cite{gal2016dropout,pop2018deep} and Bayesian neural networks \cite{rakesh2021efficacy}. Furthermore, accounting for the propagation of uncertainty from the rheological model into the continuum simulation is an important consideration for future work.

\section*{Acknowledgments}
We acknowledge helpful discussions with Joel Clemmer and Joseph Monti regarding DEM simulations in LAMMPS. This work was supported by the U.S. Department of Energy (DOE), Office of Science, Office of Advanced Scientific Computing Research, Applied Mathematics Program under contract No. DE-AC02-05CH11231. This research used resources of the National Energy Research Scientific Computing Center (NERSC), DOE Office of
Science User Facility supported by the Office of Science of the U.S. Department of Energy under Contract No. DE-AC02-05CH11231 using NERSC award ASCR-ERCAP0026881.

\section*{Data availability}
The code used to carry out the work in this manuscript is available at this \href{https://github.com/siddanib/adaptive_multiscale_granular_flows}{GitHub repository}. Our multiscale modeling code utilized the following open-source software: AMReX \cite{zhang2019amrex}, LAMMPS \cite{thompson2022lammps}, pyAMReX \cite{pyAMReX}, pyblock \cite{pyblock}, and PyTorch \cite{paszke2019pytorch}.

\appendix

\section{Additional details of multiscale approach}
\label{appendix_multiscale}

\begin{algorithm}[H]
\caption{Adaptive multiscale method}
\label{alg:multiscale_workflow}
\begin{algorithmic}[1]
\State $a \gets 0$; $b \gets 0$
\State $\bm{I}_{\text{train}} \gets \{\}$
\State $\bm{I}_{0}$ : Perform few (low $I$) DEM simulations to generate initial training data
\State $\bm{I}_{\text{train}} \gets \bm{I}_{0}$ 
\State Train $\{\mu_{m}\}_{m=1}^{M}$ using $\bm{I}_{\text{train}}$; $a \gets a + 1$
\State $I^{N} \gets \min(\bm{I}_{\text{train}})$ \Comment{low cutoff inertial number}
\While{ $t < T$} \Comment{Continuum simulation temporal loop}
    \State $\bm{I}_{\text{domain}}$ : Get $I$ for the entire domain
    \State $\bm{I}_{\text{new}} = $ {\sffamily New DEM simulations}
    
    \Comment{Presented in Algorithm \ref{alg:new_train_points}}
    
    \While{$\text{size}(\bm{I}_{\text{new}}) > 0$}
        \State Perform new DEM simulation for $\bm{I}_{\text{new}}[0]$
        
        \Comment{$\bm{I}_{\text{new}}[0]$ is the first element in $\bm{I}_{\text{new}}$}
        \State $\bm{I}_{\text{train}} \gets \bm{I}_{\text{train}} \cup \{\bm{I}_{\text{new}}[0]\}$
        \State Re-train $\{\mu_{m}\}_{m=1}^{M}$ using $\bm{I}_{\text{train}}$; $a \gets a + 1$
        \State Re-evaluate $\bm{I}_{\text{new}}$ with updated $\bm{I}_{\text{train}} \, \& \, \{\mu_{m}\}_{m=1}^{M}$
    \EndWhile
    \item[]
    \If{$a > b$} \Comment{Obtain new $I^{N}$ if the model is updated}
        \State {\sffamily Iterative refinement for} $I^{N}$ \Comment{Presented in Algorithm \ref{alg:neutral_num}}
        \State $b \gets a$
    \EndIf
    \State Obtain $\bm{\mu}_{\text{domain}}$ using $\{\mu_{m}\}_{m=1}^{M}, I^{N}, \bm{I}_{\text{domain}}$

    \Comment{Presented in Algorithm \ref{alg:mu_domain}}
    \State Update $t$
\EndWhile
\end{algorithmic}
\end{algorithm}

\begin{algorithm}[H]
\caption{New DEM simulations}
\label{alg:new_train_points}
\begin{algorithmic}[1]
\footnotesize
\Require $\bm{I}_{\text{train}}, \bm{I}_{\text{domain}}, I^{N}, \{\mu_{m}\}_{m=1}^{M}$, $a$, $b$
\State $\hat{\bm{I}}_{\text{new}} \gets $ Unique$(\bm{I}_{\text{domain}})$ \Comment{Obtain unique entries}
\State $\hat{\bm{I}}_{\text{new}} \gets \hat{\bm{I}}_{\text{new}} - \bm{I}_{\text{train}}$ \Comment{Remove training dataset points}
\item[]
\State $\bm{I}_{\text{low}} \gets \{\}$
\For{$I$ in $\hat{\bm{I}}_{\text{new}}$}
    \If{$I_{\text{DEM}}^{\min} \leq I < \min(\bm{I}_{\text{train}})$}
        \State $\bm{I}_{\text{low}} \gets \bm{I}_{\text{low}} \cup \{I\}$; 
    \EndIf
\EndFor
\State $\bm{I}_{\text{low}} \gets \quad \Uparrow\!(\bm{I}_{\text{low}})$ \Comment{Ascending sorted}
\item[] 
\State $\bm{I}_{\text{high}} \gets \{\}$
\For{$I$ in $\hat{\bm{I}}_{\text{new}}$}
    \If{$\max(\bm{I}_{\text{train}}) < I \leq I_{\text{DEM}}^{\max}$}
        \State $\bm{I}_{\text{high}} \gets \bm{I}_{\text{high}} \cup \{I\}$; 
    \EndIf
\EndFor
\State $\bm{I}_{\text{high}} \: \gets \quad \Downarrow\!(\bm{I}_{\text{high}})$ \Comment{Descending sorted}
\item[]
\State $\bm{I}_{\text{interior}} \gets \{\}$
\For{$I$ \textbf{in} $\hat{\bm{I}}_{\text{new}}$}
    \If{$\min(\bm{I}_{\text{train}}) < I < \max(\bm{I}_{\text{train}})$}
        \State Get $I_{\text{train}}^{(i)} \& I_{\text{train}}^{(i+1)} \in \bm{I}_{\text{train}}$ such that $I_{\text{train}}^{(i)} < I < I_{\text{train}}^{(i+1)}$

        \Comment{Obtain bounding training points}
        \State $\beta \gets \frac{\ln(I) - \ln(I_{\text{train}}^{(i)})}{\ln(I_{\text{train}}^{(i+1)}) - \ln(I_{\text{train}}^{(i)})}$
        \item[]
        \State $\mathcal{R} : \text{std}(\{\mu_{m}\}_{m=1}^{M}) /\text{mean}(\{\mu_{m}\}_{m=1}^{M})$
        \item[]
        \If{$\mathcal{R}(I) > \kappa \left[(1-\beta) \mathcal{R}(I_{\text{train}}^{(i)})+ \beta \mathcal{R}(I_{\text{train}}^{(i+1)})\right]$}
            \State $\bm{I}_{\text{interior}} \gets \bm{I}_{\text{interior}} \cup \{I\}$ 
        \EndIf
    \EndIf
\EndFor \Comment{$\bm{I}_{\text{interior}}$ are sorted in descending order of uncertainty}
\item[]
\State $\bm{I}_{\text{new}} \gets \bm{I}_{\text{low}} \cup \bm{I}_{\text{high}} \cup \bm{I}_{\text{interior}}$

\If{$a > b$ \&\& $\text{size}(\bm{I}_{\text{new}}) == 0$}
\State $\bm{I}_{\text{new}}$ = {\sffamily New DEM simulation to find} $I^{N}$ \Comment Presented in Algorithm \ref{alg:initial_neutral_num} 
\EndIf

\State \Return $\bm{I}_{\text{new}}$ \Comment{$\bm{I}_{\text{new}}$ can be $\{\}$}
\end{algorithmic}
\end{algorithm}

\begin{algorithm}
\caption{New DEM simulation to find $I^{N}$}
\label{alg:initial_neutral_num}
\begin{algorithmic}[1]
    \Objective Ensure that $I^{N} \in [\min(\bm{I}_{\text{train}}),\max(\bm{I}_{\text{train}}) )$  
    \Require $I^{N}, \bm{I}_{\text{train}}, \bm{I}_{\text{new}}, \{\mu_{m}\}_{m=1}^{M}$
    \State $\bm{I}_{\text{new}} \gets \{\}$
    \State $\bm{I}_{C} = \{\textbf{for} \: I \: \textbf{in} \: \bm{I}_{\text{train}} \: \textbf{if} \: \mathcal{C}(I) \leq 0\}$ \Comment{see Eq. \eqref{eq:regularization_cond}}
    \If{$\text{size}(\bm{I}_{C}) == 0$}
        \State $\bm{I}_{\text{new}} \gets \{\min(2 \max(\bm{I}_{\text{train}}),I_{\text{DEM}}^{\max})\}$
    \Else
        \State $I^{N} = \min(\bm{I}_{C})$
        \If{$I^{N} == \max(\bm{I}_{\text{train}})$ \&\& $I^{N} < I_{\text{DEM}}^{\max}$}
            \State $\bm{I}_{\text{new}} \gets \{\min(2 \max(\bm{I}_{\text{train}}),I_{\text{DEM}}^{\max})\}$
        \EndIf
    \EndIf
    \State \Return $\bm{I}_{\text{new}}$ \Comment{$\bm{I}_{\text{new}}$ can be $\{\}$}
\end{algorithmic}
\end{algorithm}

\begin{algorithm}[H]
\caption{Iterative refinement to obtain $I^{N}$}
\label{alg:neutral_num}
\begin{algorithmic}[1]
\Objective Smallest $I$ in $[\min(\bm{I}_{\text{train}}), \max(\bm{I}_{\text{train}}))$ that satisfies $\mathcal{C}(I) \leq 0$
\Require $I^{N}, \min(\bm{I}_{\text{train}}), \{\mu_{m}\}_{m=1}^{M}$
\State $\bm{I}_{C} = \{\textbf{for} \: I \: \textbf{in} \: \bm{I}_{\text{train}} \: \textbf{if} \: \mathcal{C}(I) \leq 0\}$
\State $I^{N} \gets \min(\bm{I}_{C})$
\If{$I^{N} > \min(\bm{I}_{\text{train}})$}
    \State $i \gets 0$
    \While{$i < 5$}
        \State $\bm{I}_{\text{refine}} = \, \text{linspace}(\min(\bm{I}_{\text{train}}), I^{N}, 10)$
        
        \Comment{10 linearly equidistributed points}
        \State $\bm{I}_{C} = \{\textbf{for} \: I \: \textbf{in} \: \bm{I}_{\text{refine}} \: \textbf{if} \: \mathcal{C}(I) \leq 0\}$
        \State $I^{N} \gets \min(\bm{I}_{C})$
        \State $i \gets i+1$ 
    \EndWhile
\EndIf
\State \Return $I^{N}$
\end{algorithmic}
\end{algorithm}

\begin{algorithm}[H]
\caption{Populating $\bm{\mu}_{\text{domain}}$}
\label{alg:mu_domain}
\begin{algorithmic}[1]
\Require $\{\mu_{m}\}_{m=1}^{M}, I^{N}, \bm{I}_{\text{domain}}$
\State $\mu_{1}^{N} = \left(\sum\limits_{m=1}^{M} \mu_{m}(I^{N})\right)/M$ \Comment{Ensemble mean}
\State  $A_{-} =  I^{N} \exp\left(\frac{\alpha}{\left(\mu_{1}^{N}\right)^2}\right)$ \Comment{$\alpha = 1.9$ \cite{barker2017partial}}
\For{$(\mu, I)$ \textbf{in} $(\bm{\mu}_{\text{domain}},\bm{I}_{\text{domain}})$}
    \If{$I \leq I^{N}$} \Comment{Low inertial numbers branch}
        \State $\mu(I) = \sqrt{\frac{\alpha}{\ln\left(\frac{A_{-}}{I + \epsilon_{1}}\right)}}$ \Comment{$\epsilon_{1} = 1 \times 10^{-18}$}
    \Else \Comment{Moderate and high inertial number branch}
        \State $\mu(I) = \left(\sum\limits_{m=1}^{M} \mu_{m}(I)\right)/M$
        \State $\mu(I) = \min(\sqrt{2}, \mu(I))$
    \EndIf
\EndFor
\State \Return $\bm{\mu}_{\text{domain}}$ 
\end{algorithmic}
\end{algorithm}

\section{Additional details of sub-aerial granular column collapse}

\subsection{Grid sensitivity of the continuum model}
\label{appendix_grid_sensitivity}
\begin{figure}[H]
    \centering
     \begin{subfigure}[b]{0.48\textwidth}
        \centering
        \includegraphics[width=\textwidth,keepaspectratio=true]{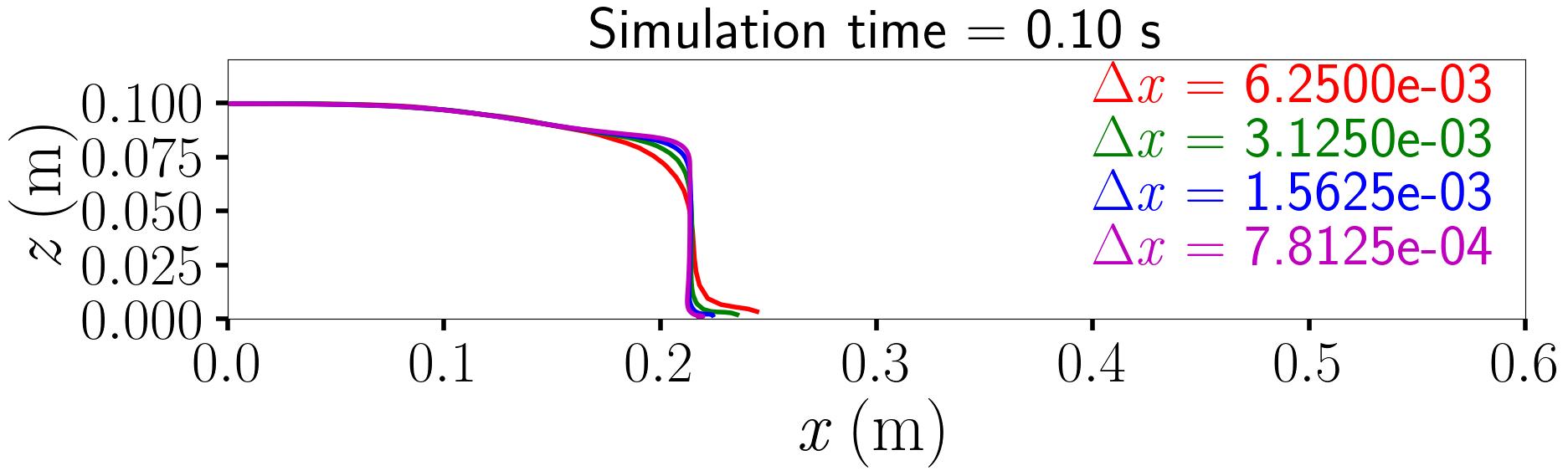}
    \end{subfigure}
    \begin{subfigure}[b]{0.48\textwidth}
        \centering
        \includegraphics[width=\textwidth,keepaspectratio=true]{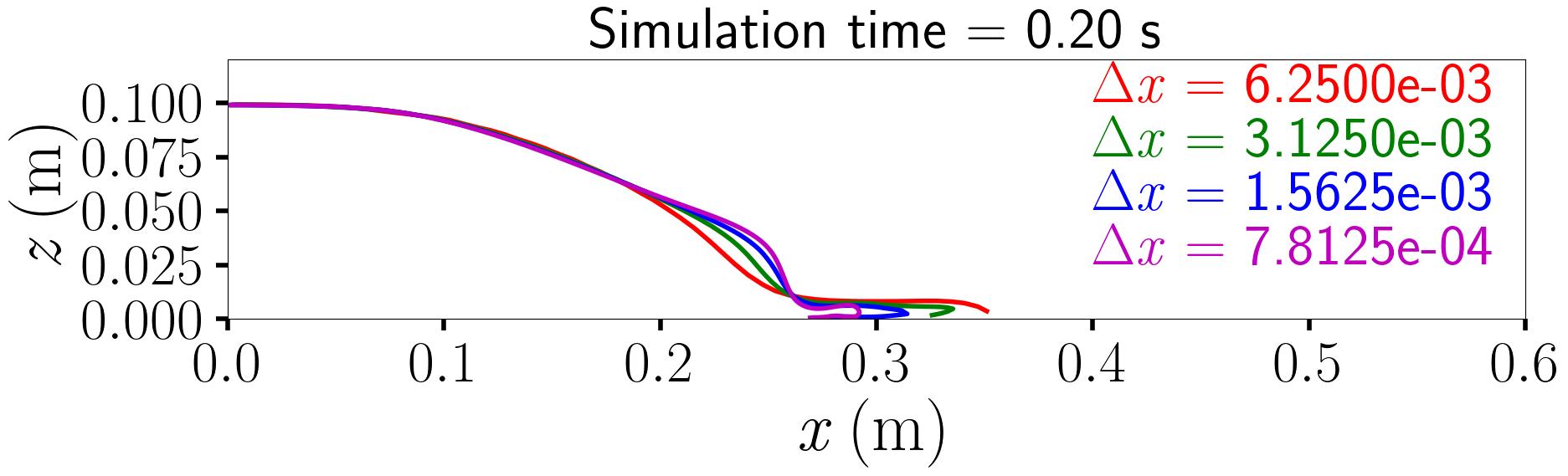}
    \end{subfigure}
    \begin{subfigure}[b]{0.48\textwidth}
        \centering
        \includegraphics[width=\textwidth,keepaspectratio=true]{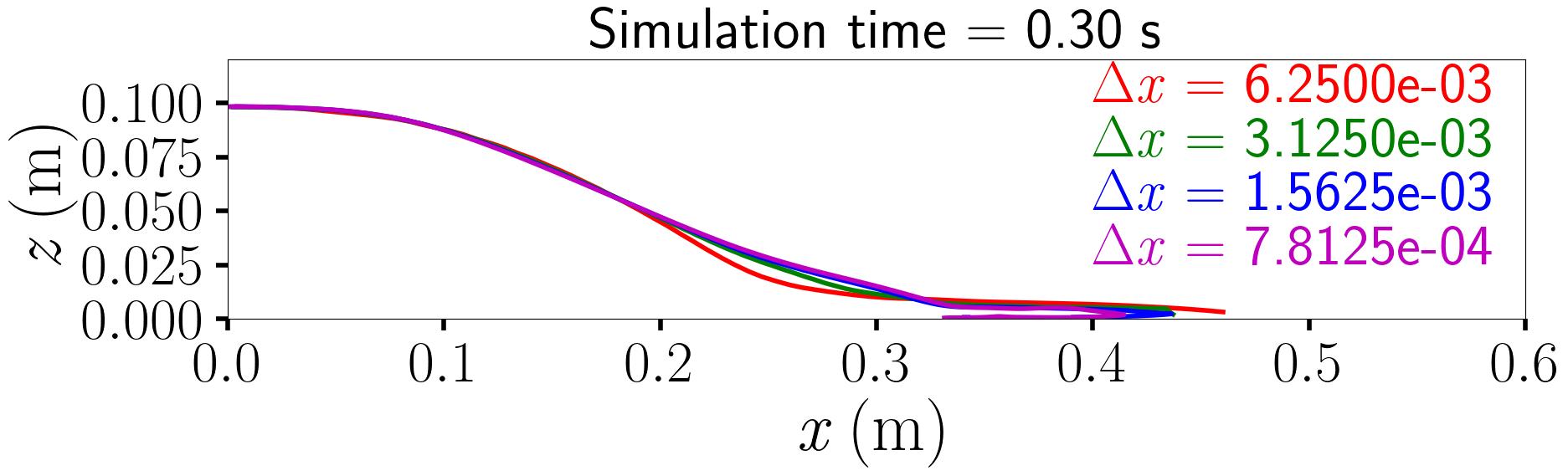}
    \end{subfigure}
    \begin{subfigure}[b]{0.48\textwidth}
        \centering
        \includegraphics[width=\textwidth,keepaspectratio=true]{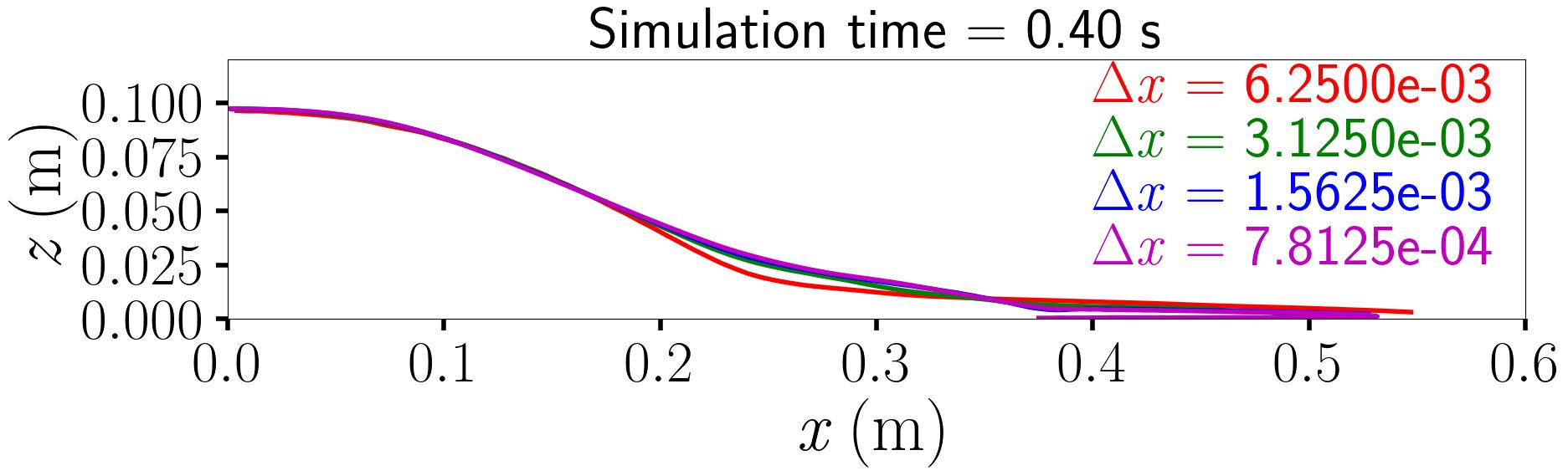}
    \end{subfigure}
    \centering
    \begin{subfigure}[b]{0.48\textwidth}
        \centering
        \includegraphics[width=\textwidth,keepaspectratio=true]{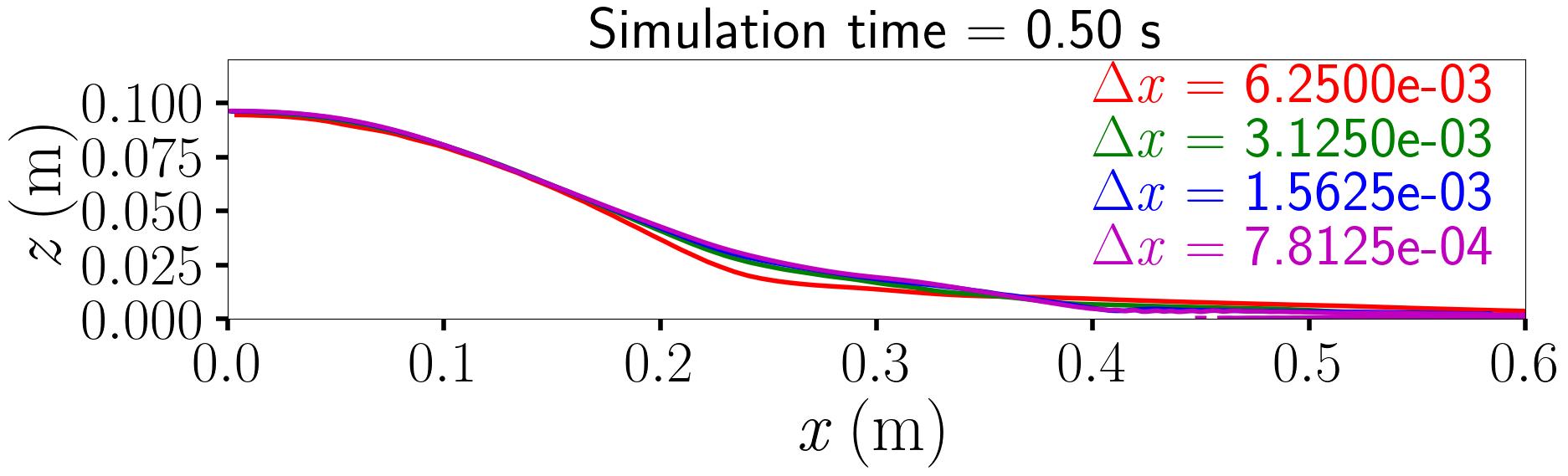}
    \end{subfigure}
    \begin{subfigure}[b]{0.48\textwidth}
        \centering
        \includegraphics[width=\textwidth,keepaspectratio=true]{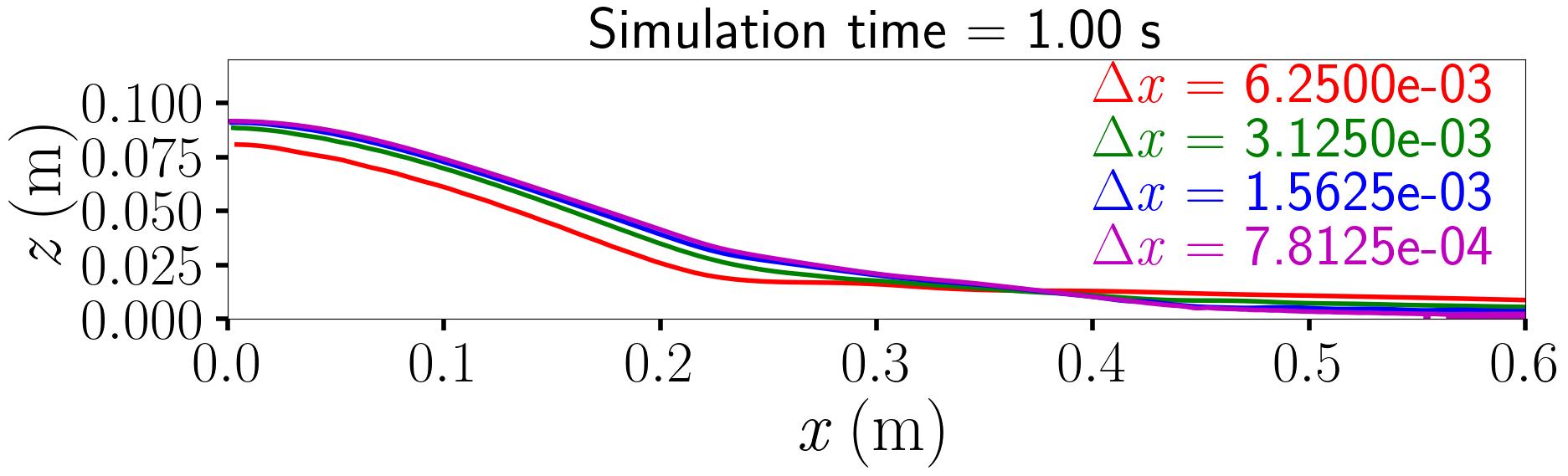}
    \end{subfigure}
    \caption{Grid sensitivity analysis for the granular column collapse setup with $H_c/L_c = 0.5$ using multiscale approach. The figure shows $c = 0.5$ contour lines at four different grid resolutions.}
    \label{fig:grid_sensitivity}
\end{figure}

The granular column collapse with $H_c/L_c=0.5$ is simulated at four levels of grid resolution, $\Delta x = \{6.25 \times 10^{-3}\text{m},3.125 \times 10^{-3}\text{m}, 1.5625 \times 10^{-3}\text{m}, 7.8125 \times 10^{-4}\text{m}\}$. The same NN-ensemble model that was adaptively learned during the course of the simulation with $\Delta x = 6.25 \times 10^{-3}$m is used as the initial model in the finer resolution simulations. The evolution of $c = 0.5$ contour line for all the considered grid resolutions at different simulation times is shown in Figure~\ref{fig:grid_sensitivity}. The contour lines of the three finer resolutions are in good agreement with each other. Therefore, the continuum simulation with $\Delta x = 1.5625 \times 10^{-3}$m is used in the current study.

\subsection{Additional details of CFD-DEM}
\label{appendix_cfd_dem}
The CFD-DEM system involves $\approx 2.3\times10^{6}$ particles with diameter uniformly distributed in the range $[0.95d_{p}, 1.05d_{p}]$. A uniform grid with $\Delta x = \Delta y = \Delta z = 6.25 \times 10^{-3}$m is used to discretize the domain. A no-slip boundary condition is imposed for the fluid on all walls. The flow solver is simulated with a constant time step of $1\times10^{-4}$s. Each time step of the flow solver involves approximately 100 DEM time steps. Further details regarding the parameters used in the simulation are reported in the Table~\ref{tab:cfd_dem_column_collapse}.
\begin{table}[H]
    \centering
    \resizebox{\columnwidth}{!}{\begin{tabular}{|l|c|}
        \hline
        Description & Value\\
        \hline
        Particle-particle friction coefficient & 0.3 \\
        Particle-wall friction coefficient & 0.317 \\
        Particle-particle coefficient of restitution & 0.9 \\
        Particle-wall coefficient of restitution & 1.0 \\
        Particle-particle normal spring constant & $2.5\times 10^{3}$Nm\textsuperscript{-1} \\
        Particle-wall normal spring constant & $2.5\times 10^{3}$Nm\textsuperscript{-1} \\
        Particle-particle tangential to normal spring constant ratio & $\approx 0.286$ \\
        Particle-wall tangential to normal spring constant ratio &  $\approx 0.286$ \\
        Particle-particle tangential to normal damping coefficient ratio  & 0.5 \\
        Particle-wall tangential to normal damping coefficient ratio & 1.0 \\
        Fluid-particle drag law & \cite{ding1990bubbling} \\
        \hline
    \end{tabular}}
    \caption{Additional parameters used to simulate the sub-aerial granular column gravitational collapse using the CFD-DEM approach. The relationships among the presented quantities are described in Musser {\it et al.} \cite{musser2022mfix} and references therein.}
    \label{tab:cfd_dem_column_collapse}
\end{table}

\subsection{Sensitivity to mixture viscosity rule}
\label{appendix_mixture_viscosity}
\begin{figure}[htbp]
    \centering
    \begin{subfigure}[b]{0.48\textwidth}
        \centering
        \includegraphics[width=\textwidth,keepaspectratio=true]{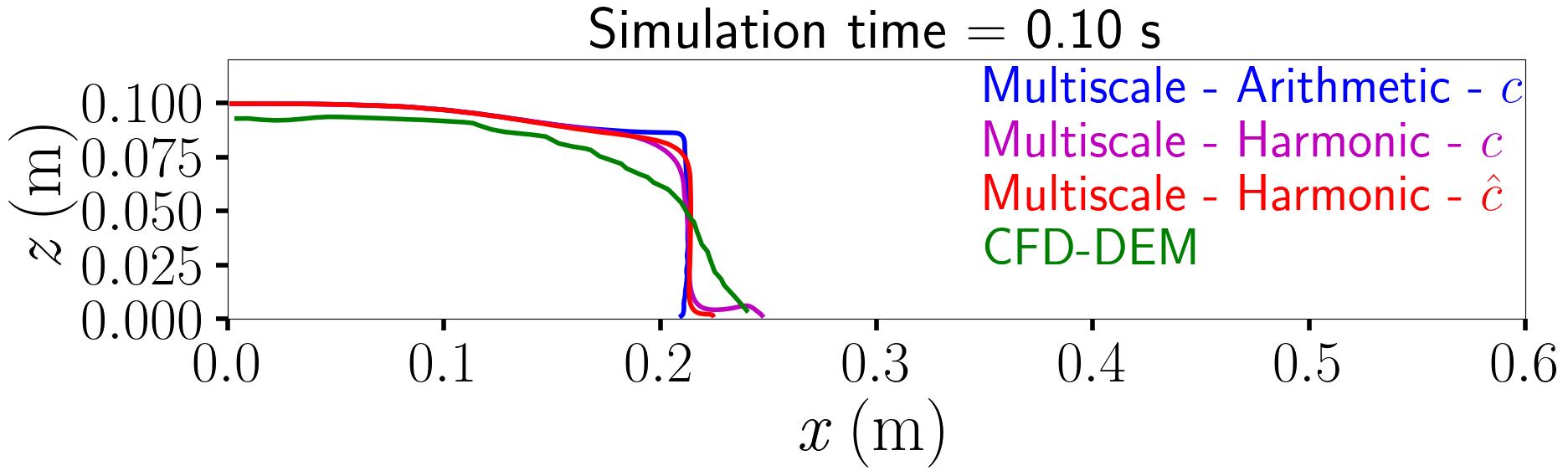}
    \end{subfigure}
    \begin{subfigure}[b]{0.48\textwidth}
        \centering
        \includegraphics[width=\textwidth,keepaspectratio=true]{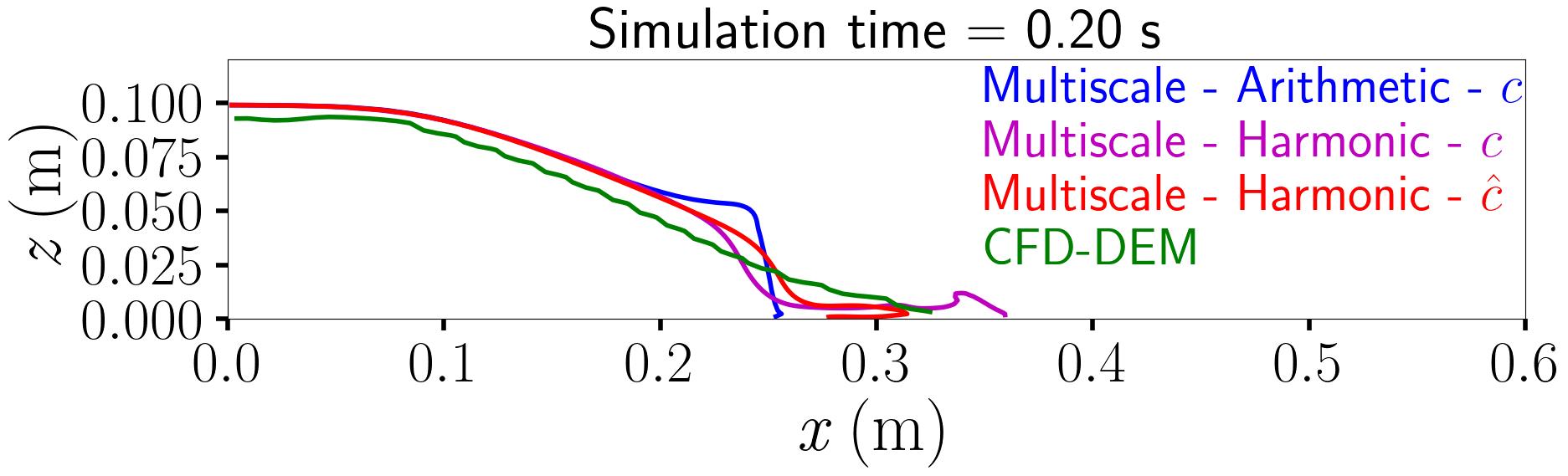}
    \end{subfigure}
    \begin{subfigure}[b]{0.48\textwidth}
        \centering
        \includegraphics[width=\textwidth,keepaspectratio=true]{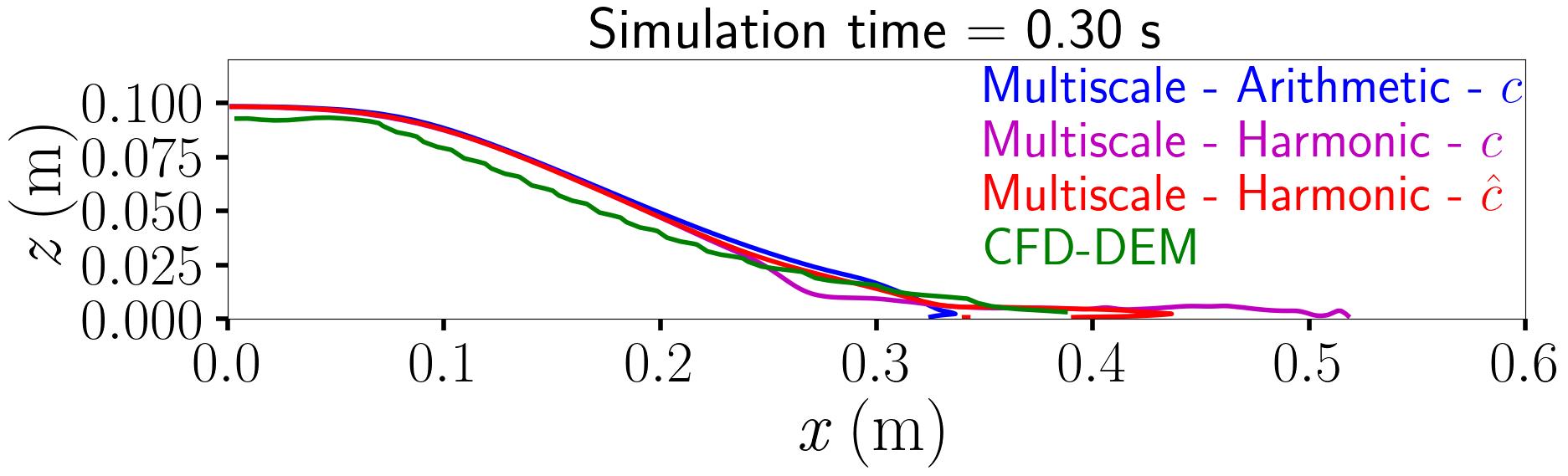}
    \end{subfigure}
    \begin{subfigure}[b]{0.48\textwidth}
        \centering
        \includegraphics[width=\textwidth,keepaspectratio=true]{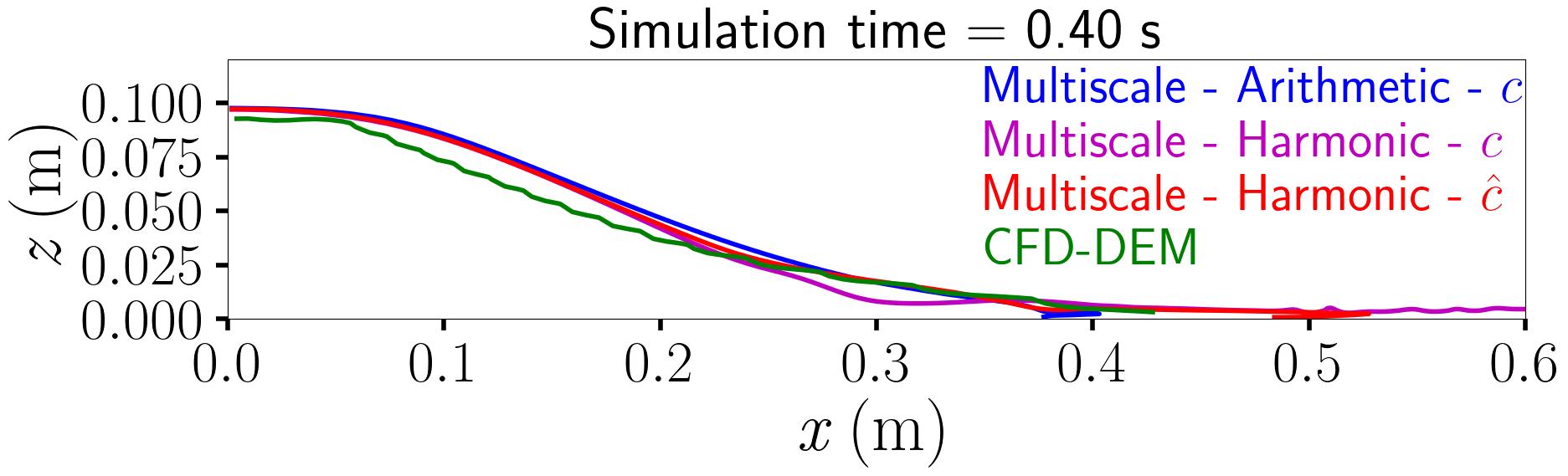}
    \end{subfigure}
    \centering
    \begin{subfigure}[b]{0.48\textwidth}
        \centering
        \includegraphics[width=\textwidth,keepaspectratio=true]{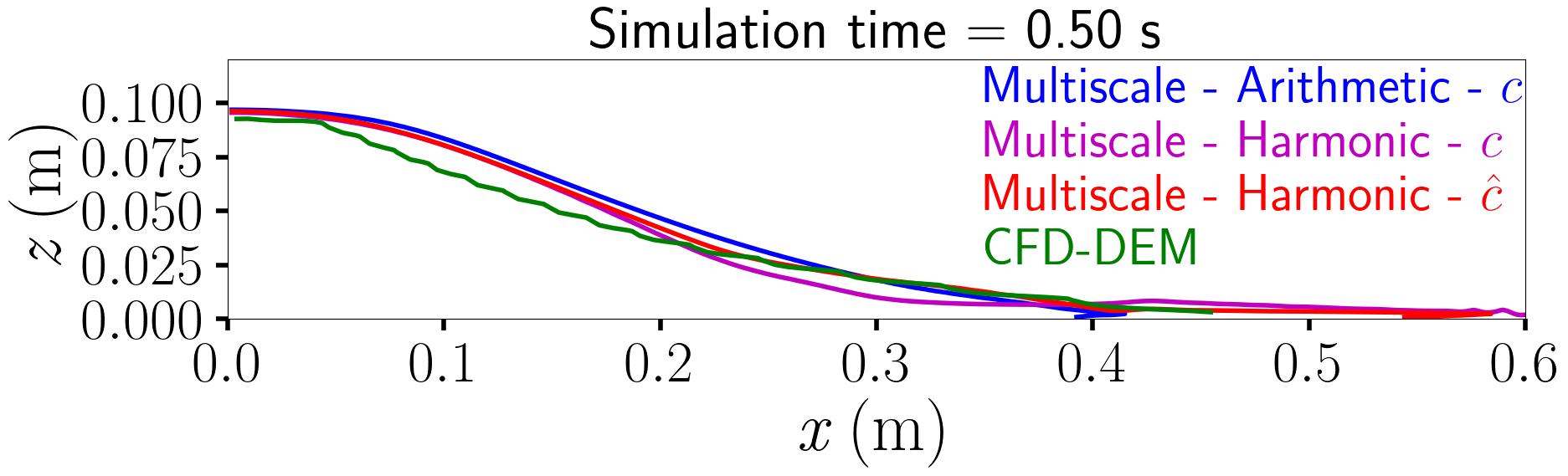}
    \end{subfigure}
    \begin{subfigure}[b]{0.48\textwidth}
        \centering
        \includegraphics[width=\textwidth,keepaspectratio=true]{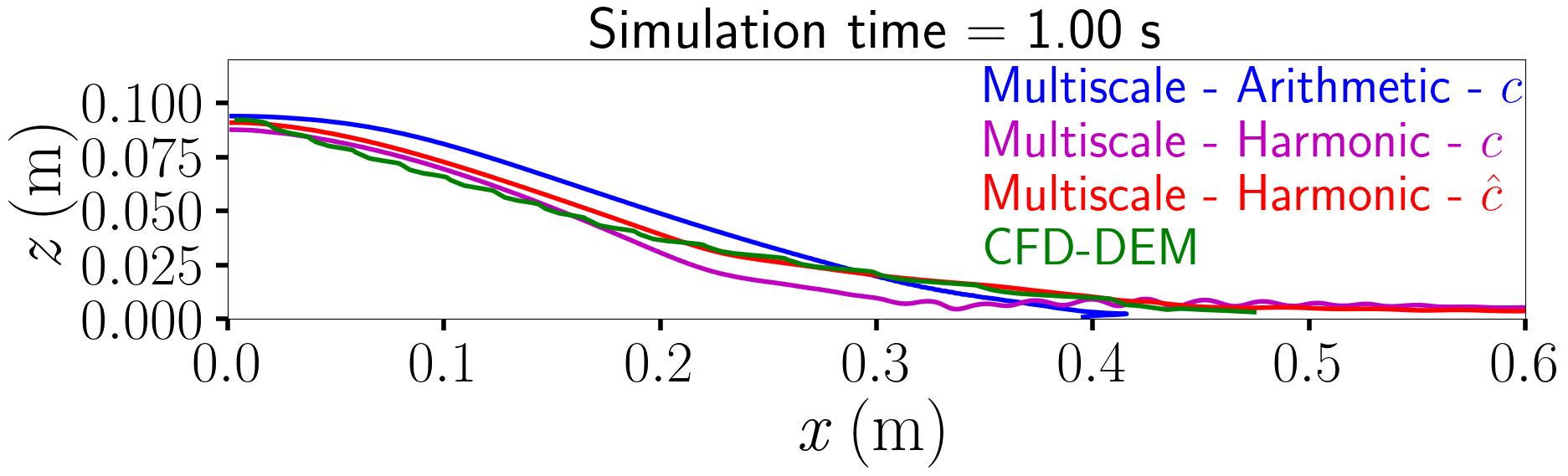}
    \end{subfigure}
    \caption{Sensitivity of the continuum model to mixture viscosity rule in the granular column collapse simulation with $H_c/L_c=0.5$. The $c = 0.5$ contour line is shown for the three continuum simulations while $\phi_{p} = 0.5$ contour line is shown for CFD-DEM simulation.}
    \label{fig:new_formulation_contour}
\end{figure}

Sensitivity of the continuum model to mixture viscosity rule is investigated by considering three different variants, which are presented below.
\begin{gather*}
    \eta^{(A)} = (1-c) \eta_f + c\eta_g\\
    \frac{1}{\eta^{(H,c)}} = \frac{1-c}{\eta_f} + \frac{c}{\eta_g}\\
    \frac{1}{\eta^{(H,\hat{c})}} = \frac{1-\hat{c}}{\eta_f} + \frac{\hat{c}}{\eta_g}; \quad  \hat{c} = \frac{\rho_g c}{\rho}
\end{gather*}
As seen from the above equations, $\eta^{(A)}$ represents the weighted arithmetic mean version, and $\eta^{(H,c)} \, \& \, \eta^{(H,\hat{c})}$ are weighted harmonic mean versions that depend on $c$ \& $\hat{c}$, respectively. The granular column collapse setup with $H_c/L_c=0.5$ is simulated using these versions at a grid resolution of $\Delta x = 1.5625 \times 10^{-3}$m. The temporal evolution of $c=0.5$ contour line for these variants along with the $\phi_{p} = 0.5$ contour line from the CFD-DEM simulation are shown in Figure~\ref{fig:new_formulation_contour}. The $\eta^{(H,\hat{c})}$ rule has the best agreement with the reference CFD-DEM simulation. Therefore, we adopt it for all the analyses presented in this work.

 \bibliographystyle{elsarticle-num}
 \bibliography{cas-refs}





\end{document}